\documentclass[a4paper, amsfonts, amssymb, amsmath, preprint, showkeys, nofootinbib, twoside]{revtex4-1}
\usepackage[english]{babel}
\usepackage[utf8]{inputenc}
\usepackage{feynmp-auto}
\usepackage{mathtools}
\usepackage{slashed}
\usepackage[colorinlistoftodos, color=green!40, prependcaption]{todonotes}
\usepackage{amsthm}
\usepackage{mathtools}
\usepackage{physics}
\usepackage{xcolor}
\usepackage{graphicx}
\usepackage[left=23mm,right=13mm,top=35mm,columnsep=15pt]{geometry} 
\usepackage{adjustbox}
\usepackage{placeins}
\usepackage[T1]{fontenc}
\usepackage{lipsum}
\usepackage{csquotes}
\usepackage[pdftex, pdftitle={Article}, pdfauthor={Author}]{hyperref} 

\newcommand{\lag}{\mathcal{L}}
\newcommand{\M}{\mathcal{M}}

\begin{document}
\title{On the Impact of Dark Matter Scattering\\ on the Trajectory of High-Energy Cosmic Rays}

\author{Stefano Profumo}
    \email[Correspondence email address: ]{profumo@ucsc.edu}
    \affiliation{Department of Physics, University of California, Santa Cruz (UCSC),
Santa Cruz, CA 95064, USA}
\affiliation{Santa Cruz Institute for Particle Physics (SCIPP),
Santa Cruz, CA 95064, USA}

\author{M. Grant Roberts}
\email{migrober@ucsc.edu}
\affiliation{Department of Physics, University of California, Santa Cruz (UCSC),
Santa Cruz, CA 95064, USA}
\affiliation{Santa Cruz Institute for Particle Physics (SCIPP),
Santa Cruz, CA 95064, USA}

\author{Shashank Dharanibalan}
\email{sdharani@ucsc.edu }
\affiliation{Department of Physics, University of California, Santa Cruz (UCSC),
Santa Cruz, CA 95064, USA}

\date{\today} 

\begin{abstract}
We study the impact on the trajectory of high-energy cosmic-ray protons of scattering off the cosmic dark matter. We compute the scattering angle as a function of the cosmic-ray energy, of the dark matter mass, and of the interaction strength for a few representative choices for the relevant interaction cross section. We find that the typical deflection angle over the cosmic ray path is largely independent of the dark matter mass. Given existing limits on the interaction strength, we compute the average deflection angle. We find that for large interaction cross sections and low cosmic ray energies, the predicted deflection angle is much larger than the angular resolution of very high-energy cosmic-ray observatories such as Pierre Auger.
\end{abstract}


\maketitle

\section{Introduction} \label{sec:outline}
The particle nature and the interaction properties of the cosmic dark matter (DM) with Standard Model fields are unknown. General constraints can be set on the largest-possible super-Planckian mass of macroscopic DM candidates from considerations such as tidal heating of structure, leading to an upper limit around stellar masses (for a recent study, see e.g. \cite{Koulen:2024emg}); at the opposite end of the spectrum, where the occupation number of DM is too high to be compatible with it being fermionic, light, bosonic DM is bound to have a mass larger than that corresponding to a de Broglie wavelength on the order of the size of the smallest observed virialized structures, dwarf galaxies \cite{Hui:2016ltb}: such limit hovers around $10^{-21}$ eV \cite{Hui:2021tkt}.

The interaction of the DM with ordinary matter, such as nucleons and electrons, is constrained at varying levels and with different methods and experiments as a function of the DM mass. Broadly, the most stringent constraints exist in the mass range where low-background, solid state, or noble gas direct detection detectors are sensitive to DM-nucleon scattering; at larger masses, constraints stem from distortions of the cosmic microwave background, as well as from so-called paleodetectors (for a recent study see e.g. \cite{Acevedo:2021tbl}). In the opposite, light-mass regime, on the other hand, constraints arise both from the so-called Migdal effect (the energy losses stemming from bremsstrahlung and other emission seeded by DM scattering with atomic nuclei) \cite{Ibe:2017yqa}, and from up-scattering of DM by cosmic rays to energies where it could be detected by neutrino detectors \cite{Cappiello:2019qsw}.

Here, we study the effect on the arrival direction of very high-energy cosmic rays (CRs) of scattering off of the cosmological DM intervening between the source and the detector. We consider both fermionic and bosonic DM candidates, and an exhaustive set of Lorentz structures that could describe the structure of the DM-nucleon interaction, as well as the possibility that a light mediator exists and thus of a ``simplified model'' for the DM-nucleon interaction as opposed to an effective operator. We then compute the typical, most probable, cumulative scattering angle for a given DM mass, interaction cross section, CR energy, and source distance.

The question of the origin of very high energy CRs hinges on the ability to utilize the arrival direction of the detected CR with its source \cite{Kachelriess:2019oqu}. While the effect of magnetic fields is certainly critical, and has been widely studied (see e.g. \cite{Garcia:2021cgu} and references therein), it is in our view equally important to assess whether scattering of CRs off of the intervening DM can also have an effect, and if so how large.

A key ingredient in this calculation is the column density of DM traversed by the CRs, defined as \cite{Granelli:2022ysi}: 
\begin{equation}
    \Sigma\equiv \int_{\rm l.o.s.} \rho_{\rm DM}(s)\ ds,
\end{equation}
where $\rho_{\rm DM}$ indicates the DM density and $ds$ is the infinitesimal length element along the line of sight (l.o.s.). There are three components to $\Sigma$: the column density $\Sigma_{\rm src}$ within the CR source host galaxy/galaxy cluster; the average column density over cosmological density, $\Sigma_{\rm cosmo}$ and, finally, the column density within the Milky Way, $\Sigma_{\rm MW}$. Ref.~\cite{Granelli:2022ysi} estimates that for selected blazars such as BL Lacertae and TXS 0506+056 
\begin{equation}
    10^{26}\lesssim\frac{\Sigma_{\rm src}/m_{\rm DM}}{{\rm cm}^{-2}}\lesssim 10^{33};
\end{equation} we estimate 
\begin{equation}
   \frac{\Sigma_{\rm cosmo}/m_{\rm DM}}{{\rm cm}^{-2}}\lesssim \frac{d_{\rm cosmo}\ \rho_{\rm DM}}{m_{\rm DM}}\sim  10^{21}
\end{equation} and 
\begin{equation}    
    \frac{\Sigma_{\rm MW}/m_{\rm DM}}{{\rm cm}^{-2}}\lesssim \frac{d_{\rm MW}\ \rho_{\rm DM}}{m_{\rm DM}}\sim  10^{22},
\end{equation}
 where we took $d_{\rm cosmo}\simeq 1$ Gpc and $d_{\rm MW}\sim 10$ kpc. Thus, the column density in the CR source host galaxy dominates the column density by anywhere between 4 and 11 orders of magnitude compared to the similar values of the column density over cosmological distances and that of the Milky Way.

 \begin{figure}[!t]
    \centering
    \includegraphics[scale=0.95]{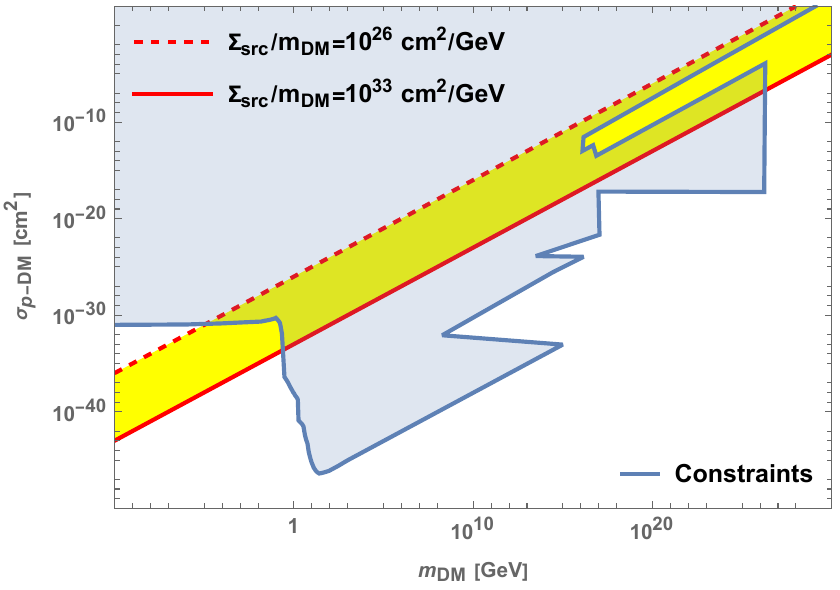}
    \caption{The blue region shows the area of the $(m_{\rm DM},\sigma_{\rm p-DM})$ parameter space excluded by a variety of constraints and experiments. The shaded yellow region shows the region of parameter space corresponding to one scattering, for values of $\Sigma_{\rm src}/m_{\rm DM}$ in the range we consider plausible; regions above those lines have multiple scattering events, on average.}
    \label{fig:intro}
\end{figure}

Fig.~\ref{fig:intro} displays, on the  $(m_{\rm DM},\sigma_{\rm p-DM})$ parameter space (where $m_{\rm DM}$ is the DM mass, and $\sigma_{\rm p-DM}$ the DM-proton interaction cross section), the constraints from a variety of experiments \cite{Cappiello:2019qsw, Clark:2020mna} with blue shading. The red solid line corresponds to $\Sigma \sigma_{\rm p-DM}=1$ for ${\Sigma_{\rm src}/m_{\rm DM}}\lesssim 10^{33}\ {\rm cm}^{-2}$, while the red-dashed line corresponds to ${\Sigma_{\rm src}/m_{\rm DM}}\lesssim 10^{26}\ {\rm cm}^{-2}$. The yellow region highlights the parameter space where one scattering is, on average, expected within the plausible range of column density values. Above those lines, multiple scattering events are possible. The figure thus highlights that only in the very light and in the very heavy mass ranges is the scattering of high-energy CR off of DM statistically possible. We will thus focus on those regions in what follows.

Our results detailed in the remainder of this study indicate that in most cases there exists an {\em absolute upper limit} to the compounded scattering angle in CR-DM scattering; depending on the assumed column density and interaction cross section, the angular resolution of very high-energy CR (VHECR) telescopes, such as the Pierre Auger Observatory \cite{PierreAuger:2015eyc}.

The remainder of the manuscript is organized as follows. First, we discuss the kinematics of the process under consideration in sec.~\ref{sec:kinematics}; we then outline and detail in sec.~\ref{sec:operators} the relevant effective operators and simplified models we will consider in our studies; we then provide the results of the cross section, and classify the leading kinematic dependence of different classes of operators in sec.~\ref{sec:xsec}; Sec.~\ref{sec:results} presents our results; the final sec.~\ref{sec:conclusions} presents our discussion, outlook, and conclusions.

\section{Kinematics} \label{sec:kinematics}
We follow here the discussion in Ref.~\cite{Dick_2009} in deriving the angular distribution in the ``laboratory frame'' given a cross section. We indicate with $\theta$ is the scattering angle in the center of mass frame and $\theta^*$ is the scattering angle in the laboratory frame. 

\vspace{1 cm}
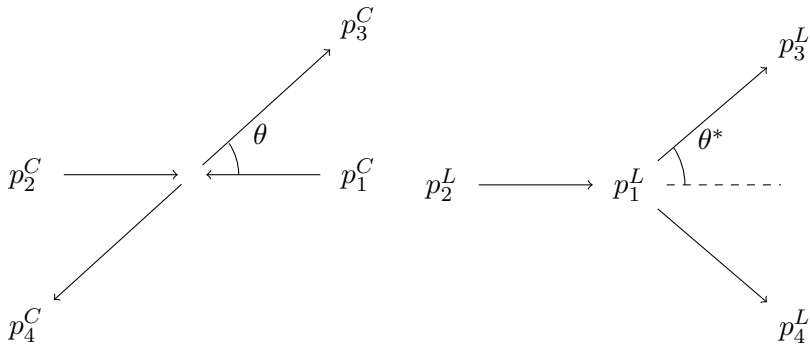
\begin{figure}[h]
\centering
\begin{tikzpicture}[
roundnode/.style={circle, minimum size=0mm},
squarednode/.style={rectangle, minimum size=7mm},
]
\node[roundnode] (2)                          {$p_2^C$};
\node[roundnode] (c)      [right=15mm of 2]   {};
\node[roundnode] (1)      [right=15mm of c]   {$p_1^C$};
\node[roundnode] (3)      [above=of 1]        {$p_3^C$};
\node[roundnode] (4)      [below=of 2]        {$p_4^C$};
\node[roundnode] (temp)   [left=6mm of 1]     {};
\node[roundnode] (a*)     [above=0.1mm of temp]  {$\theta$};
\draw[->]   (2) -- (c);
\draw[->]   (1) -- (c);
\draw[->]   (c) -- (3);
\draw[->]   (c) -- (4);
\draw (28mm,0mm) arc (0:33.69:7.5mm);
\end{tikzpicture} 
\begin{tikzpicture}[
roundnode/.style={circle, minimum size=0mm},
squarednode/.style={rectangle, minimum size=7mm},
]
\node[roundnode] (2)                         {$p_2^L$};
\node[roundnode] (c)   [right=15mm of 2]     {$p_1^L$};
\node[roundnode] (1)   [right=15mm of c]     {};
\node[roundnode] (3)   [above=12mm of 1]     {$p_3^L$};
\node[roundnode] (4)   [below=12mm of 1]     {$p_4^L$};
\node[roundnode] (temp)[right=4mm of c]      {};
\node[roundnode] (a)   [above=0.1mm of temp] {$\theta^*$};
\draw[->]      (2) -- (c);
\draw[->]      (c) -- (3);
\draw[->]      (c) -- (4);
\draw[dashed]  (c) -- (1);
\draw (32mm,0mm) arc (0:33.69:9mm);
\end{tikzpicture}
\caption{The center of mass frame (left) and laboratory frame (right) kinematics.}
\end{figure}
%
%
%
The four momenta in the laboratory frame in the initial state are $p^L_1=(m_\chi,0)$ for the dark matter, which we thus assume to be initially at rest, and $p^L_2=(E_2,\vec p_2)$ for the impinging cosmic ray. The Mandelstam variables $s$, $t$, and $u$, which we utilize to express the cross section, can be cast as:
\begin{equation}
    s=(m_\chi+E_2)^2-(E_2^2-m_N^2),
\end{equation}
\begin{equation}
    t=2m_\chi(E_3-E_2),
\end{equation}
\begin{equation}
    u=2m_\chi^2+2m_M^2-s-t,
\end{equation}
 where $m_N$ is the cosmic-ray proton mass, and where $E_3$ is the final-state cosmic-ray energy; the latter can be expressed as follows:
\begin{equation}
    a=\frac{s+m_N^2-m_\chi^2}{2}
\end{equation}
\begin{equation}
    b=E_2+m_\chi
\end{equation}
\begin{equation}
    D=b^2-(E_2^2-m_N^2)\cos\theta^*
\end{equation}
\begin{equation}
    E_{3,\pm}=\frac{ab}{D}\pm\frac{\sqrt{E_2^2-m_N^2}\cos\theta^*}{D}\sqrt{a^2-m_N^2\Big[b^2-\left(E_2^2-m_N^2\right)\cos^2\theta^*\Big]}.
\end{equation}
Notice that there exist two values of $E_3=E_{3,+},\ E_{3,-}$ for every value of $E_2$, and that a maximal scattering angle exists as long as $m_\chi<m_N$, with a value
\begin{equation}
    \theta^*_{\rm max}={\rm arccos}\left(1-\frac{m^2_\chi}{m^2_N}\right).
\end{equation}
Defining 
\begin{equation}
    \lambda_{\chi N}=\left(s-m_\chi^2-m_N^2\right)^2-4m_N^2m_\chi^2,
\end{equation}
it is possible to relate the differential cross section in the center of mass frame, $d\sigma/d\Omega$ to the cross section in the laboratory frame, $d\sigma/d\Omega^*$ as follows
\begin{equation}
    \frac{d\sigma}{d\Omega^*}=\sum_{\pm}\frac{4m_\chi s}{\lambda_{\chi N}}\frac{|\vec p^L_{1}||\vec p^L_{3,\pm}|}{E_2+m_\chi-\frac{|\vec p^L_{1}|}{|\vec p^L_{3,\pm}|}E_{3,\pm}\cos\theta^*} \frac{d\sigma}{d\Omega}(s,t_\pm,u_\pm),
\end{equation}
with 
\begin{equation}
    |\vec p^L_{1}|=\sqrt{E_2^2-m_N^2}
\end{equation}
and \begin{equation}
    |\vec p^L_{3,\pm}|=\sqrt{E_{3,\pm}^2-m_N^2}.
\end{equation}
The expression above is what we utilize in the remainder of the paper to analyze the scattering angle for CR-DM collisions.

\section{Operators}\label{sec:operators}
We assume that the DM is a Dirac spin 1/2 fermionic particle $\chi$, and we consider the following Lorentz structure for the case where the mediator of the nucleon-DM interactions is heavy, and the interaction can be described in terms of an effective dimension-6 4-fermion operator (hereafter we indicate the nucleon with $N$):
\begin{itemize}
\item {\bf Scalar operators:} 
\begin{itemize}
    \item $\bar\chi\chi\bar N N$
    \item $\bar\chi\gamma^5\chi\bar N N$ 
    \item $\bar\chi\chi\bar N\gamma^5 N$
    \item $\bar\chi\gamma^5\chi\bar N\gamma^5 N$
\end{itemize}

\item {\bf Vector operators:}
\begin{itemize}
    \item $\bar\chi\gamma^\mu\chi\bar N\gamma_\mu N$ 
    \item $\bar\chi\gamma^5\gamma^\mu\chi\bar N\gamma_\mu N$ 
    \item $\bar\chi\gamma^\mu\chi\bar N\gamma_\mu\gamma^5 N$
    \item $\bar\chi\gamma^5\gamma^\mu\chi\bar N\gamma_\mu\gamma^5 N$ 
\end{itemize}
\item {\bf Tensor operators:}
\begin{itemize}
    \item $\bar\chi\sigma^{\mu\nu}\gamma^5\chi\bar N\sigma_{\mu\nu} N$
    \item $\bar\chi\sigma^{\mu\nu}\chi\bar N\sigma_{\mu\nu} N$
\end{itemize}
\end{itemize}
If the mediator mass is comparable or smaller than the center of mass energy of the process, the effective theory description is invalid. In that regime, we consider, for simplicity, the case of a scalar mediator, for which the general interaction Lagrangian can be cast as
\begin{equation}
    \lag_{\rm{int}} = i\kappa\left(\Bar{\chi}\Gamma_{a} \chi + \Bar{N}\Gamma_{b} N \right)\phi,
\end{equation}

\noindent where $\Gamma_{a}$ and $\Gamma_{b}$ are the generic bilinears listed above for the effective operators. The relevant diagram is then:\\
\begin{center}
\begin{fmffile}{second-diagram}
 \begin{fmfgraph*}(120,80)
  \fmfleft{i1,i2}
  \fmfright{o1,o2}
  \fmf{fermion,label=$N$}{i2,v1}
  \fmf{fermion,label=$\chi$}{i1,v2}
  \fmf{dashes,label=$\phi$}{v1,v2}
  \fmf{fermion,label=$N$}{v1,o2}
  \fmf{fermion,label=$\chi$}{v2,o1}
  \fmflabel{$i\kappa$}{v1}
  \fmflabel{$i\kappa$}{v2}
 \end{fmfgraph*}
\end{fmffile}
\end{center}


Because of the assumption of a scalar mediator, the change with respect to the operators considered above is very simple, as one simply picks up a factor of $\frac{\kappa^{2}}{q^{2} - m_{\phi}^{2}} = \frac{\kappa^{2}}{t - m_{\phi}^{2}}$, where $t$ is the Mandelstam variable defined above. So we can simply modify all the amplitudes from the previous sections as:

\begin{equation}
    |\M|^{2} = \frac{\kappa^{4}}{(t - m_{\phi}^{2})^{2}}|\M_{o}|^{2}.
\end{equation}

\noindent In other words, we simply just scale each of the previous $|\M|^{2}$ by the new factor and replace $g{^2}$ with $\kappa^{4}$. This also allows us to calculate $g$ in terms of $\kappa$ and $m_{\phi}$, 
\begin{equation*}
    \frac{\kappa^{4}}{(t - m_{\phi}^{2})^{2}} = \frac{\kappa^{4}}{m_{\phi}^{4}(1 - m_{\phi}^{2}/t)^{2}} \sim \frac{\kappa^{4}}{m_{\phi}^{4} } + \dots
\end{equation*}

\noindent As a result, $g^{2} = {\kappa^{4}}/{m_{\phi}^{4}}$.

 \section{Cross Sections}\label{sec:xsec}
In what follows we utilize the following naming convention, where for brevity the symbol $\chi$ indicates the DM particle:

\begin{center}
    \begin{fmffile}{first-diagram}
 \begin{fmfgraph*}(120,80)
  \fmfleft{i1,i2}
  \fmfright{o1,o2}
  \fmf{fermion,label=$N$}{i2,v1}
  \fmf{fermion,label=$\chi$}{i1,v1}
  \fmf{fermion,label=$N$}{v1,o2}
  \fmf{fermion,label=$\chi$}{v1,o1}
  \fmflabel{$ig$}{v1}
 \end{fmfgraph*}
\end{fmffile}

\noindent 1 = initial state $\chi$; 2 = initial state $N$; 3 = final state N; 4 = final state $\chi$
\end{center}

\subsection{$\bar\chi\chi\bar N N$}

\noindent The scattering amplitude is given by:

\begin{equation}
    \M = ig[\Bar{u}(4)u(1)][\Bar{u}(3)u(2)].
\label{eq:invariant-amplitude}
\end{equation}

\noindent To calculate the amplitude squared, we average as usual over initial spins and sum over final spins. By using Casimir's identities, we have:

\begin{equation}
    \sum_{\text{all spins}}[\Bar{u}(a)\Gamma_{1} u(b)][\Bar{u}(a)\Gamma_{2} u(b)]^{*} = Tr[\Gamma_{1}(\slashed{p}_{b} + m_{b})\Bar{\Gamma}_{2}(\slashed{p}_{a} + m_{a})].
\end{equation}


\noindent For the amplitude under investigation we have $\Gamma_{1} = \Gamma_{2} = 1$. Applying the result of Eq.~\ref{eq:invariant-amplitude}, we find

\begin{equation*}
    |\M|^{2} = g^{2}\left[ 4(p_{1}\cdot p_{4}) + 4m_{\chi}^{2} \right] \left[ 4(p_{2}\cdot p_{3}) + 4m_{N}^{2} \right],
\end{equation*}

\begin{equation}
    |\M|^{2} = \frac{1}{4}16g^{2}\left[ (p_{1}\cdot p_{4})(p_{2}\cdot p_{3}) + m_{N}^{2}(p_{1}\cdot p_{4}) + m_{\chi}^{2}(p_{2}\cdot p_{3}) + (m_{\chi}m_{N})^{2} \right].
\label{eq:invariant-amplitude-spin-average}
\end{equation}

\noindent Notice that the factor of $\frac{1}{4}$ above comes averaging over the initial spins. Generically, the cross-section is given by:

\begin{equation}
    d\sigma = \frac{1}{4I}|\M|^{2}d\Phi^{(n)},
\label{eq:general-cross-section}
\end{equation}

\noindent where $I = \sqrt{(p_{1}\cdot p_{2})^{2} - (m_{1}m_{2})^{2}}$ is the flux factor, and $d\Phi^{(n)}$ is given by,

\begin{equation}
    d\Phi^{(n)} = (2\pi)^{4}\delta^{4}(p_{\text{initial}} - p_{\text{final}})\prod_{i=1}^{n} \frac{d^{3}p_{i}}{(2\pi)^{3}(2E_i)},
\end{equation}

\noindent where $n$ is the number of particles in the final state. For our case, $n = 2$, thus $d\Phi^{(2)}$ is,

\begin{equation}
    d\Phi^{(2)} = (2\pi)^{4}\delta^{4}(p_{1} + p_{2} - p_{3}-p_{4})\frac{d^{3}p_{3}}{(2\pi)^{3}2E(p_{3})}\frac{d^{3}p_{4}}{(2\pi)^{3}2E(p_{4})}
\end{equation}

\noindent Using the 3-momentum delta function, we can set $p_{4} = p_{1} + p_{2} - p_{3}$, which eliminates the $d^{3}p_{4}$ integration. We are then left with $\delta(E_{1} + E_{2} - E_{3} - E_{4}(E_{3})$ and an integral over $p_{3}$. Using the relativistic dispersion relation, $E^{2} = p^{2} + m^{2}$ we can convert the $d^{3}p = p_{3}^{2}dp$ integral into a $dE_{3}$ integral. Explicitly,

\begin{align}
    &p_{1} = (E_{1}, \Vec{p}_{1}), ~~~p_{2} = (E_{2}, -\Vec{p}_{1}),\\
    &p_{4} = (E_{4}, \Vec{p}_{4}), ~~~p_{3} = (E_{3}, -\Vec{p}_{4})
\end{align}

\noindent We also now specify our frame of reference to be the center of mass frame, where $\Vec{p}_{2} = -\Vec{p}_{1}$ and $\Vec{p}_{3} = -\Vec{p}_{4}$, which then explicitly implies that

\begin{equation}
    E_{4}^{2} = E_{3}^{2} - m_{N}^{2} + m_{\chi}^{2}.
\label{eq:E4}
\end{equation}

\noindent Using the delta function identity:

\begin{equation*}
    \delta(g(x)) = \frac{\delta(x-x_{0})}{|\frac{dg(x_{0})}{dx}|},
\end{equation*}

\noindent where $x_{0}$ is the root of $g(x)$. Setting $g(E_{3}) = E_{1} + E_{2} - E_{3} - E_{4}(E_{3})$, the root ($E_{30}$) is given by:

\begin{equation}
    E_{30} = \frac{E^{2}+\Delta}{2E} = \frac{(E_{1} + E_{2})^{2} + m_{N}^{2} - m_{\chi}^{2}}{2(E_{1} + E_{2})}.
\label{eq:E30}
\end{equation}

\noindent Carrying out the energy integral we get,

\begin{equation}
    d\Phi^{(2)} = \frac{1}{4(2\pi)^{2}}\frac{\sqrt{E_{30}^{2} - m_{N}^{2}}}{E_{40} + E_{30}}d\Omega,
\end{equation}

\noindent because $E_{4}$ is a function of $E_{3}$, when using the new energy delta function, we define $E_{40} = E_{4}(E_{30})$. We now write the scattering amplitude in terms of the energies and the scattering angle:

\begin{align*}
    &(p_{1}\cdot p_{4})(p_{2}\cdot p_{3}) = E_{1}E_{40}E_{2}E_{30} + \sqrt{E_{1}^{2}-m_{\chi}^{2}} \sqrt{E_{30}^{2} - m_{N}^{2}} \cos(\theta)(E_{1}E_{40} + E_{2}E_{30}) + \\& \qquad\qquad\qquad\qquad (E_{1}^{2} - m_{\chi}^{2})(E_{30}^{2} - m_{N}^{2})\cos^{2}(\theta),\\
    &m_{N}^{2}(p_{1}\cdot p_{4}) = m_{N}^{2}E_{1}E_{40} + m_{N}^{2}|\Vec{p_{1}}||\Vec{p_{3}}|\cos(\theta) = m_{N}^{2}E_{1}E_{40} + m_{N}^{2}\sqrt{E_{1}^{2}-m_{\chi}^{2}}\sqrt{E_{30}^{2} - m_{N}^{2}}\cos(\theta),\\
    &m_{\chi}^{2}(p_{2}\cdot p_{3}) = m_{\chi}^{2}E_{2}E_{30} + m_{\chi}^{2}|\Vec{p_{1}}||\Vec{p_{3}}|\cos(\theta) = m_{\chi}^{2}E_{2}E_{30} + m_{\chi}^{2}\sqrt{E_{1}^{2}-m_{\chi}^{2}}\sqrt{E_{30}^{2} - m_{N}^{2}}\cos(\theta).\\
\end{align*}





\noindent Thus, the scattering amplitude becomes,

\begin{align}
    |\M|^{2} = \frac{1}{4}16g^{2} [E_{1}E_{40}E_{2}E_{30} + (E_{1}^{2} - m_{\chi}^{2})(E_{30}^{2} - m_{\chi}^{2})\cos^{2}(\theta) + m_{N}^{2}E_{1}E_{40} + m_{\chi}^{2}E_{2}E_{30} + \\
    \left(m_{N}^{2} + m_{\chi}^{2} + E_{1}E_{40} + E_{2}E_{30}\right)\sqrt{E_{1}^{2}-m_{\chi}^{2}}\sqrt{E_{30}^{2} - m_{N}^{2}}\cos(\theta)
    (m_{\chi}m_{N})^{2}]\nonumber .
\end{align}

\noindent The flux factor, $I$, is defined in terms of $(p_{1}\cdot p_{2})^{2} = \left(E_{1}E_{2} + E_{1}^{2} - m_{\chi}^{2}\right)^{2}$, thus giving:

\begin{equation*}
    I = \sqrt{\left(E_{1}(E_{1} + E_{2}) - m_{\chi}^{2}\right)^{2} - m_{\chi}^{2}m_{N}^{2}}.
\end{equation*}

\noindent The differential cross-section is then, finally,

\begin{multline}
    \frac{d\sigma}{d\Omega} = \frac{g^{2}}{16\pi^{2}}\frac{1}{\sqrt{\left(E_{1}(E_{1} + E_{2}) - m_{\chi}^{2}\right)^{2} - m_{\chi}^{2}m_{N}^{2}}}\frac{\sqrt{E_{30}^{2} - m_{N}^{2}}}{E_{40} + E_{30}}\\ \times [ E_{1}E_{40}E_{2}E_{30} + (E_{1}^{2} - m_{\chi}^{2})(E_{30}^{2} - m_{\chi}^{2})\cos^{2}(\theta) + m_{N}^{2}E_{1}E_{40} + m_{\chi}^{2}E_{2}E_{30} + \\ \times
    \left(m_{N}^{2} + m_{\chi}^{2} + E_{1}E_{40} + E_{2}E_{30}\right)\sqrt{E_{1}^{2}-m_{\chi}^{2}}\sqrt{E_{30}^{2} - m_{N}^{2}}\cos(\theta)
    +(m_{\chi}m_{N})^{2} ].
\label{eq:center-of-mass-cross-section}
\end{multline}



We now wish to express the cross section in terms of the Mandelstam

\begin{align}
    &s = (p_{1}+p_{2})^{2} = (p_{3}+p_{4})^{2},\\
    &t = (p_{1}-p_{4})^{2} = (p_{3}-p_{2})^{2},\\
    &u = (p_{1}-p_{3})^{2} = (p_{4}-p_{2})^{2}.
\end{align}

\noindent In addition, note that, in the center of mass frame, $\sqrt{s} = E_{1} + E_{2}$. The relevant scalar products are thus:

\begin{align}
    &p_{1}\cdot p_{4} = \frac{1}{2}(2m_{\chi}^{2} - t),\\
    &p_{2}\cdot p_{3} = \frac{1}{2}(2m_{N}^{2} - t),\\
    &p_{1}\cdot p_{2} = \frac{1}{2}(s-m_{\chi}^{2} - m_{N}^{2}).
\end{align}

\noindent With the expressions above, we can rewrite Eq.~\ref{eq:center-of-mass-cross-section} in terms of $t$ and $s$,

\begin{multline}
    \frac{d\sigma}{d\Omega^{*}} = \frac{g^{2}}{16\pi^{2}}\frac{1}{\sqrt{(s-m_{\chi}^{2} - m_{N}^{2})^{2} - 4m_{\chi}^{2}m_{N}^{2}}}  \frac{\sqrt{(s + m_{N}^{2} - m_{\chi}^{2})^{2} - 4m_{N}^{2}s}}{s} \\ \times \left[ \frac{1}{4}t^{2} - t\left(m_{\chi}^{2}+m_{N}^{2}\right) +  4m_{\chi}^{2}m_{N}^{2} \right].
\end{multline}

\noindent Note that 

\begin{align}
    E_{30} = \frac{(s + \Delta)}{2\sqrt{s}} = \frac{(s + m_{N}^{2} - m_{\chi}^{2})}{2\sqrt{s}},\\
    E_{40} = \frac{(s - \Delta)}{2\sqrt{s}} = \frac{(s - m_{N}^{2} + m_{\chi}^{2})}{2\sqrt{s}}
\end{align}

\noindent We can further simplify the differential cross-section further by multiplying out the terms in the square roots; this reduces the differential cross-section to the following, much simpler form:

\begin{equation}
    \frac{d\sigma}{d\Omega^{*}} = \frac{g^{2}}{16\pi^{2}}\frac{1}{s}\left[ \frac{1}{4}t^{2} - t\left(m_{\chi}^{2}+m_{N}^{2}\right) +  4m_{\chi}^{2}m_{N}^{2} \right].
\end{equation}

\noindent There are thus two contributions from the combined $s$ and $t$ channels, and one term contributing solely to the $s$ channel. In the limit of small momentum transfer, the cross-section goes like $1/s$, which is enhanced at slow velocity, and decays quickly for large velocity interactions. For non-zero momentum transfer, we have equally likely contributions of forward and backwards scattering. As a side note, this calculation tells us that for a generic center of mass scattering, we can write,

\begin{equation}
    \frac{d\sigma}{d\Omega^{*}} = \frac{g^{2}}{64\pi^{2}}\frac{1}{s}|\M|^{2}.
\end{equation}

\subsection{$\Bar{\chi}\gamma^{\mu}\gamma^{5}\chi\Bar{N}\gamma_{\mu}\gamma^{5}N$}

\noindent We now consider the case of the following interaction Lagrangian,

\begin{equation}
    \lag_{\rm{int}} = ig\Bar{\chi}\gamma^{\mu}\gamma^{5}\chi\Bar{N}\gamma_{\mu}\gamma^{5}N.
\end{equation}

\noindent The invariant amplitude is then of the form,

\begin{equation}
    \M_{PVPV} = ig[\Bar{u}(4)\gamma^{\mu}\gamma^{5}u(1)][\Bar{u}(3)\gamma_{\mu}\gamma^{5}u(2)].
\end{equation}

\noindent Applying Casimir's trick as before, we find that for each trace operation,

\begin{equation*}
    Tr[...] = 4\left(-g^{\mu\nu}(p_{1} \cdot p_{4}) + p_{1}^{\nu}p_{4}^{\mu} + p_{1}^{\mu}p_{4}^{\nu}\right) - 4m_{\chi}^{2}g^{\mu\nu},
\end{equation*}

\noindent which means that the invariant amplitude squared is given by,

\begin{equation}
    |\M|^{2}_{PVPV} = \frac{1}{4}32g^{2}\left[(p_{1} \cdot p_{3})(p_{2} \cdot p_{4}) + (p_{1} \cdot p_{2})(p_{3} \cdot p_{4}) + m_{N}^{2}(p_{1} \cdot p_{4}) + m_{\chi}^{2}(p_{2} \cdot p_{3}) + 2m_{\chi}^{2}m_{N}^{2}\right].
\label{eq:pseudovector-pseudovector-invariant-amplitude-squared}
\end{equation}

\noindent Using our Mandelstam variable definitions, we find that,

\begin{equation}
    |\M|^{2}_{PVPV} = \frac{1}{4}32g^{2}\left[\frac{1}{4}(s^{2}+u^{2}) - \frac{1}{2}(m_{\chi}^{2}-m_{N}^{2})^{2}\right].
\end{equation}

\noindent The differential cross-section then becomes:

\begin{equation}
    \frac{d\sigma}{d\Omega^{*}} = \frac{g^{2}}{8\pi^{2}}\frac{1}{s}\left[ \frac{1}{4}(s^{2}+u^{2}) - \frac{1}{2}(m_{\chi}^{2}-m_{N}^{2})^{2} \right].
\end{equation}

\subsection{$\Bar{\chi}\gamma^{5}\chi\Bar{N}\gamma^{5}N$}

We now consider the case of the following interaction Lagrangian,

\begin{equation}
    \lag_{\rm{int}} = ig\Bar{\chi}\gamma^{5}\chi\Bar{N}\gamma^{5}N.
\end{equation}

\noindent The invariant amplitude is then of the form,

\begin{equation}
    \M_{PSPS} = ig[\Bar{u}(4)\gamma^{5}u(1)][\Bar{u}(3)\gamma^{5}u(2)].
\end{equation}

\noindent Applying Casimir's identity as before, we find that for each trace operation,

\begin{equation*}
    Tr[...] = 4(p_{a}\cdot p_{b}) - 4 m_{a}m_{b},
\end{equation*}

\noindent which means that the invariant amplitude squared will be given by,

\begin{equation}
    |\M|^{2}_{PSPS} = \frac{1}{4}16g^{2}[(p_{1} \cdot p_{4}) - m_{\chi}^{2}][(p_{2} \cdot p_{3}) - m_{N}^{2}].
\label{eq:pseudoscalar-pseudoscalar-amplitude-squared}
\end{equation}

\noindent Using  Mandelstam variables, we find that

\begin{equation}
    |\M|^{2}_{PSPS} = \frac{1}{4}4g^{2}t^{2} = g^{2}t^{2}.
\end{equation}

\noindent The differential cross-section then becomes:

\begin{equation}
    \frac{d\sigma}{d\Omega^{*}} = \frac{g^{2}}{64\pi^{2}}\frac{t^{2}}{s}
\end{equation}

\subsection{$\Bar{\chi}\gamma^{\mu}\chi\Bar{N}\gamma_{\mu}\gamma^{5}N$}

We now consider the case of the following interaction Lagrangian,

\begin{equation}
    \lag_{\rm{int}} = ig\Bar{\chi}\gamma^{\mu}\chi\Bar{N}\gamma_{\mu}\gamma^{5}N.
\end{equation}

\noindent The invariant amplitude is of the form,

\begin{equation}
    \M_{V \times AV} = ig[\Bar{u}(4)\gamma^{\mu}u(1)][\Bar{u}(3)\gamma_{\mu}\gamma^{5}u(2)].
\end{equation}

\noindent Applying Casimir's trick as before, we find that for the vector trace we have,

\begin{equation*}
    Tr[...] = 4\left[g^{\mu\nu}(m_{\chi}^{2} - (p_{1}\cdot p_{4})) + p_{1}^{\nu}p_{4}^{\mu} + p_{1}^{\mu}p_{4}^{\nu}\right].
\end{equation*}

\noindent The axial vector trace was done previously. This means that the invariant amplitude squared will be given by,

\begin{equation}
    |\M|^{2}_{V \times AV} = \frac{1}{4}32g^{2}\left[(p_{1} \cdot p_{3})(p_{2} \cdot p_{4}) + (p_{1} \cdot p_{2})(p_{3} \cdot p_{4}) + m_{N}^{2}(p_{1} \cdot p_{4}) - m_{\chi}^{2}(p_{2} \cdot p_{3}) - 2m_{\chi}^{2}m_{N}^{2}\right].
\label{eq:pseudoscalar-pseudoscalar-amplitude-squared}
\end{equation}

\noindent Using  Mandelstam variables, we find that

\begin{equation}
    |\M|^{2}_{V \times AV} = 8g^{2}\left[\frac{1}{4}(s^{2} + u^{2}) - \frac{1}{2}m_{\chi}^{2}(s - t + u) - \frac{1}{2}m_{N}^{2}(s + t + u) + \frac{1}{2}(m_{\chi}^{2} + m_{N}^{2})^{2}\right]
\end{equation}

\noindent The differential cross-section then becomes:



\begin{equation}
    \frac{d\sigma}{d\Omega^{*}} = \frac{g^{2}}{8\pi^{2}}\frac{1}{s}\left[\frac{1}{4}(s^{2} + u^{2}) - \frac{1}{2}m_{\chi}^{2}(s - t + u) + \frac{1}{2}(m_{\chi}^{2} + m_{N}^{2})(m_{\chi}^{2} - m_{N}^{2})\right],
\end{equation}

\noindent and further simplifying using $s - t + u = s + t + u - 2t = 2(m_{\chi}^{2} + m_{N}^{2}) - 2t$, we find the simplified form:

\begin{equation}
    \frac{d\sigma}{d\Omega^{*}} = \frac{g^{2}}{8\pi^{2}}\frac{1}{s}\left[\frac{1}{4}(s^{2} + u^{2}) + m_{\chi}^{2}t - \frac{1}{2}(m_{\chi}^{2} + m_{N}^{2})^{2}\right]
\end{equation}

\subsection{$\Bar{\chi}\gamma^{5}\gamma^{\mu}\chi\Bar{N}\gamma_{\mu}N$}

We now consider the case of the following interaction Lagrangian,

\begin{equation}
    \lag_{\rm{int}} = ig\Bar{\chi}\gamma^{5}\gamma^{\mu}\chi\Bar{N}\gamma_{\mu}N.
\end{equation}

\begin{equation}
    \M_{PV\times V} = ig[\Bar{u}(4)\gamma_{5}\gamma^{\mu}u(1)][\Bar{u}(3)\gamma_{\mu}u(2)].
\end{equation}

\begin{equation}
|\M|^{2}_{PV\times V} = 8g^{2}\left[(p_{1} \cdot p_{3})(p_{2} \cdot p_{4}) + (p_{1} \cdot p_{2})(p_{3} \cdot p_{4}) - m_{N}^{2}(p_{1} \cdot p_{4}) + m_{\chi}^{2}(p_{2} \cdot p_{3}) - 2m_{\chi}^{2}m_{N}^{2}\right].
\label{eq:vector-vector-amplitude-squared}
\end{equation}

\noindent Using Mandelstam variables, we find that,

\begin{equation}
    |\M|^{2}_{PV \times V} = 2g^{2}(s^{2} + u^{2}).
\end{equation}

\subsection{$\Bar{\chi}\chi\Bar{N}\gamma^{5}N$}

We now consider the case of the following interaction Lagrangian,

\begin{equation}
    \lag_{\rm{int}} = ig\Bar{\chi}\chi\Bar{N}\gamma^{5}N.
\end{equation}

\begin{equation}
    \M_{SPS} = ig[\Bar{u}(4)u(1)][\Bar{u}(3)\gamma^{5}u(2)].
\end{equation}

\begin{equation}
    |\M|^{2}_{SPS} = 4g^{2}[(p_{1} \cdot p_{4}) + m_{\chi}^{2}][(p_{2} \cdot p_{3}) - m_{N}^{2}].
\label{eq:pseudoscalar-scalar-amplitude-squared}
\end{equation}

\noindent Using the Mandelstam definitions, we find that

\begin{equation}
    |\M|^{2}_{SPS} = g^{2}t^{2}.
\end{equation}

\subsection{$\bar{\chi}\sigma^{\mu\nu}\chi\Bar{N}\sigma_{\mu\nu}N$}

We now consider the case of the following interaction Lagrangian,

\begin{equation}
    \lag_{\rm{int}} = ig\Bar{\chi}\sigma^{\mu\nu}\chi\Bar{N}\sigma_{\mu\nu}N.
\end{equation}

\begin{equation}
    \M_{TT} = ig[\Bar{u}(4)\sigma^{\mu\nu}u(1)][\Bar{u}(3)\sigma_{\mu\nu}u(2)].
\end{equation}

\begin{equation}
|\M|^{2}_{TT} = 32g^{2}\left[-(p_{1} \cdot p_{4})(p_{2} \cdot p_{3}) + 2(p_{1} \cdot p_{3})(p_{2} \cdot p_{4}) + 2(p_{1} \cdot p_{2})(p_{3} \cdot p_{4}) + 3m_{\chi}^{2}m_{N}^{2}\right].
\label{eq:vector-vector-amplitude-squared}
\end{equation}

\noindent Using Mandelstam variables, we find that,

\begin{equation}
    |\M|^{2}_{TT} = 16g^{2}(s^{2} + u^{2}).
\end{equation}



\subsection{Leading order behavior of the remaining operators}

We will now simplify the process and simply list out the relevant operator, its invariant amplitude, the square of the amplitude to leading order, and the corresponding cross-section to leading order. 

\subsubsection{$\Bar{\chi}\gamma^{5}\chi\Bar{N}N$}

We now consider the case of the following interaction Lagrangian,

\begin{equation}
    \lag_{\rm{int}} = ig\Bar{\chi}\gamma^{5}\chi\Bar{N}N.
\end{equation}

\begin{equation}
    \M_{PSS} = ig[\Bar{u}(4)\gamma^{5}u(1)][\Bar{u}(3)u(2)].
\end{equation}

\begin{equation}
    |\M|^{2}_{PSS} = \frac{1}{4}16g^{2}[(p_{1} \cdot p_{4}) - m_{\chi}^{2}][(p_{2} \cdot p_{3}) - m_{N}^{2}].
\label{eq:pseudoscalar-scalar-amplitude-squared}
\end{equation}

\noindent Using Mandelstam variables, we find that,

\begin{equation}
    |\M|^{2}_{PSS} = g^{2}t^{2}.
\end{equation}

\subsubsection{$\Bar{\chi}\gamma^{\mu}\chi\Bar{N}\gamma_{\mu}N$}

We now consider the case of the following interaction Lagrangian,

\begin{equation}
    \lag_{\rm{int}} = ig\Bar{\chi}\gamma^{\mu}\chi\Bar{N}\gamma_{\mu}N.
\end{equation}

\begin{equation}
    \M_{V\times V} = ig[\Bar{u}(4)\gamma^{\mu}u(1)][\Bar{u}(3)\gamma_{\mu}u(2)].
\end{equation}

\begin{equation}
|\M|^{2}_{V\times V} = 8g^{2}\left[(p_{1} \cdot p_{3})(p_{2} \cdot p_{4}) + (p_{1} \cdot p_{2})(p_{3} \cdot p_{4}) - m_{N}^{2}(p_{1} \cdot p_{4}) - m_{\chi}^{2}(p_{2} \cdot p_{3}) + 2m_{\chi}^{2}m_{N}^{2}\right].
\label{eq:vector-vector-amplitude-squared}
\end{equation}

\noindent Using Mandelstam variables, we find that

\begin{equation}
    |\M|^{2}_{V \times V} = 2g^{2}(s^{2} + u^{2}).
\end{equation}

\section{Results}\label{sec:results}
Using the results of the computation of the cross sections detailed in the previous section, we now compute the differential cross section for CR-DM interaction as a function of the scattering angle in the ``laboratory'' frame. This will inform us of the most-likely angle that the scattering process produces. We will argue that all ``scalar'', ``vector'', and the ``tensor'' Lorentz structures display, up to overall rescaling factors, the same behavior in the relevant angular distribution. 

\begin{table}[]
    \centering
    \begin{tabular}{|c|c|c|}
    \hline
      & $\bar\chi\chi\bar N N$  & $t^2$ \\
& $\bar\chi\gamma^5\chi\bar N N$  & $t^{2}$     \\
SI &  $\bar\chi\gamma^\mu\chi\bar N\gamma_\mu N$  & $s^{2}+u^{2}$    \\
&  $\bar\chi\gamma^5\gamma^\mu\chi\bar N\gamma_\mu N$  & $s^{2} + u^{2}$    \\
&   $\bar\chi\sigma^{\mu\nu}\gamma^5\chi\bar N\sigma_{\mu\nu} N$  & $s^{2} + u^{2}$    \\
\hline
      &   $\bar\chi\chi\bar N\gamma^5 N$  & $t^{2}$    \\
&   $\bar\chi\gamma^5\chi\bar N\gamma^5 N$  & $t^2$  \\
SD &   $\bar\chi\gamma^\mu\chi\bar N\gamma_\mu\gamma^5 N$  & $s^2+u^2$ \\
&   $\bar\chi\gamma^5\gamma^\mu\chi\bar N\gamma_\mu\gamma^5 N$  & $s^2+u^2$  \\
 &   $\bar\chi\sigma^{\mu\nu}\chi\bar N\sigma_{\mu\nu} N$   & $s^{2}+u^{2}-t^{2}$  \\
 \hline
    \end{tabular}
    \caption{Leading terms in the Mandelstam variables for different Lorentz structures}
    \label{tab:ops}
\end{table}

The key difference between the three different ``classes'' of Lorentz structures stems from the leading terms in the high-energy limit (thus those that are mass-independent). Specifically, we find three structures:
\begin{itemize}
    \item $|{\cal M}|^2\sim t^2$ for the scalar case
    \item $|{\cal M}|^2\sim s^2+u^2$ for the vector case
    \item $|{\cal M}|^2\sim s^2+u^2-t^2$ for the tensor case    
\end{itemize}

We show the complete set of leading terms in $|{\cal M}|^2$ in Tab.~\ref{tab:ops}, where we also detail which of the cross section are, in the low-enspin-dependentin independent (SI) or spin dependent (SD).

\subsection{Scalar Lorentz structures}
\begin{figure}
    \centering
    \includegraphics[scale=0.3]{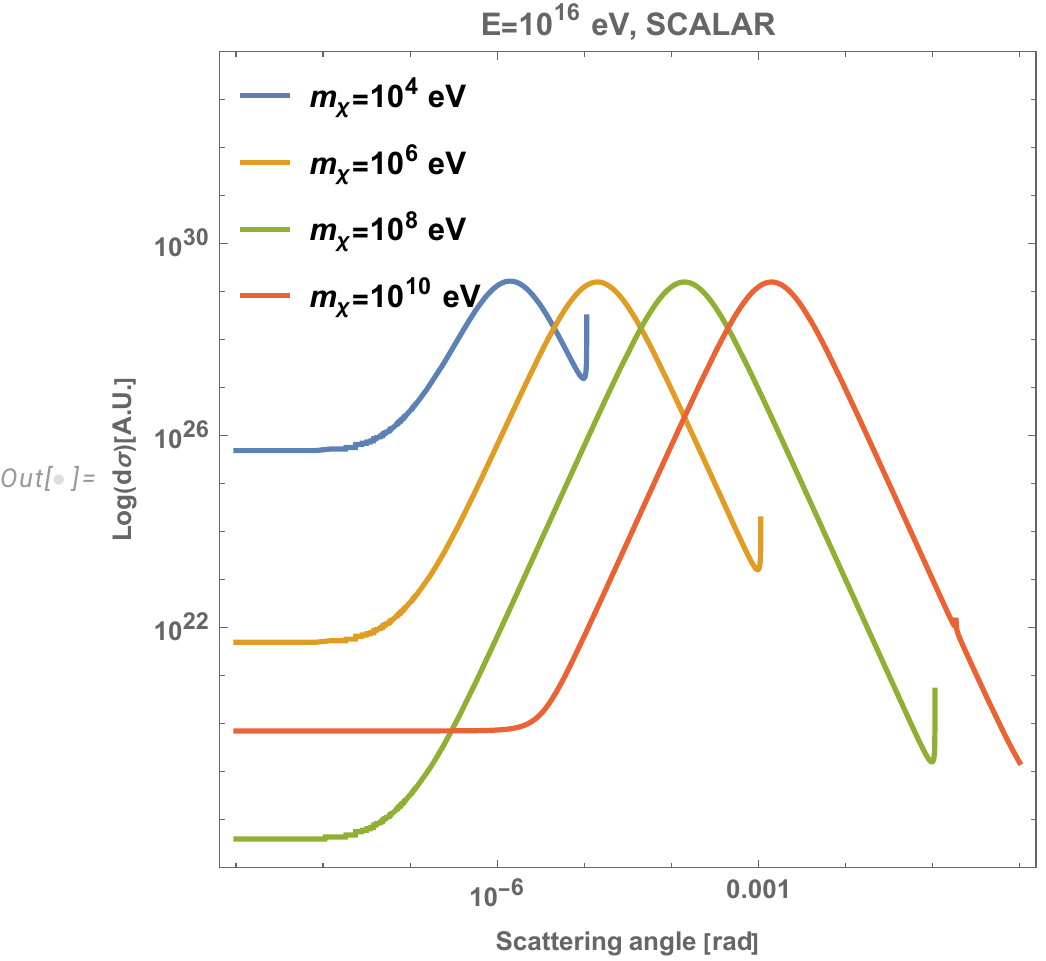}\ \ \includegraphics[scale=0.3]{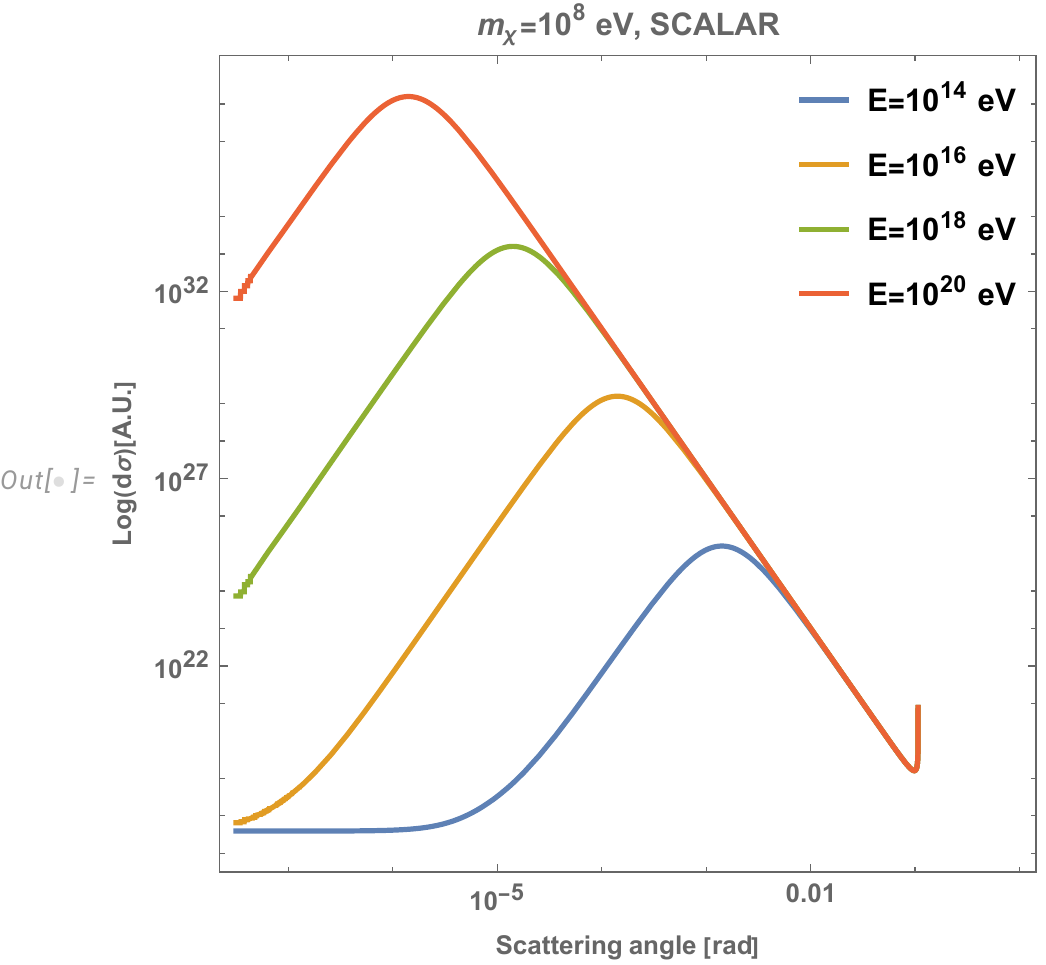}\ \ \includegraphics[scale=0.3]{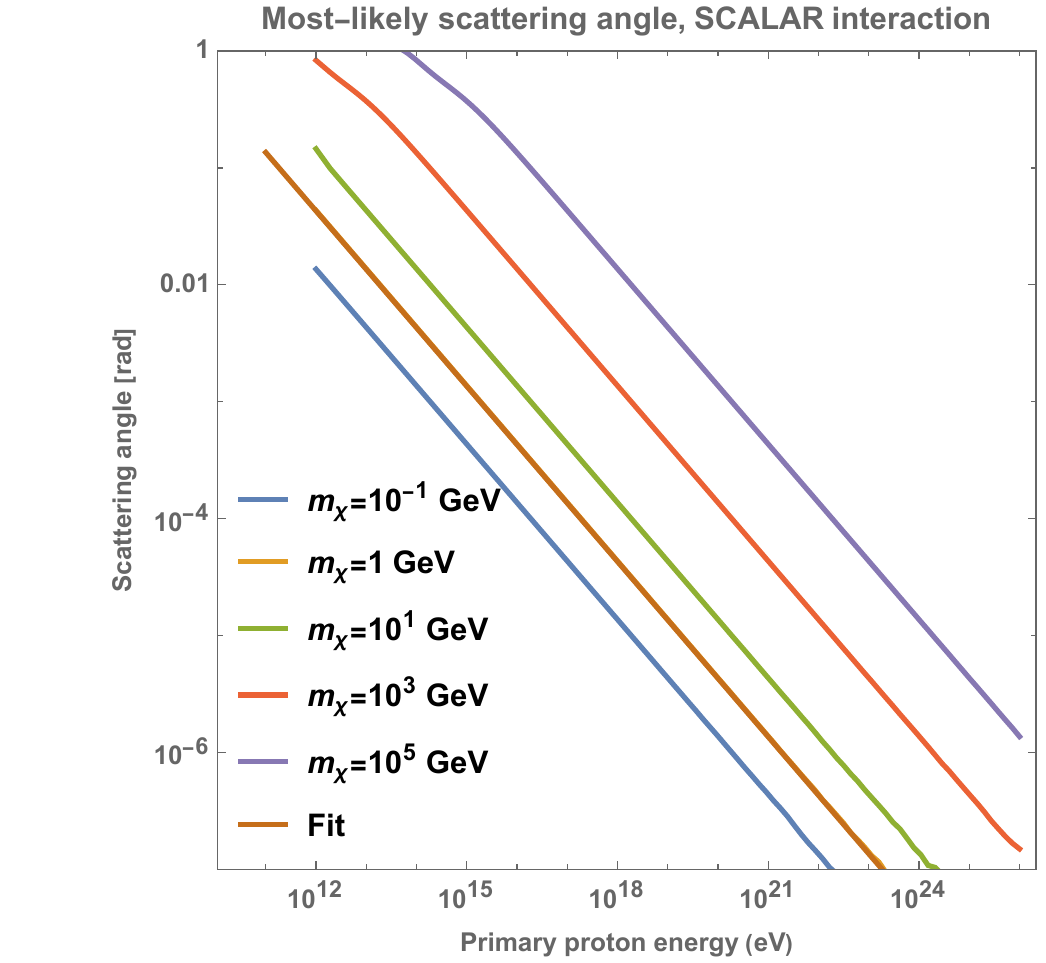}
    \caption{The product of the laboratory cross section and the laboratory scattering angle $\frac{d\sigma}{d\Omega^*}d\theta^*$, as a function of the scattering angle $\theta^*$, for the scalar-scalar cross section, at fixed energy and for different dark matter masses (left), and at fixed dark matter mass, and for different energies (center). The right panel shows the most-likely scattering angle as a function of the primary CR proton energy, for a variety of DM masses, and one instance of the fitting function shown in Eq.~(\ref{eq:thetaSS})}
    \label{fig:SS}
\end{figure}

As remarked above, all Lorentz structures listed in sec.~\ref{sec:xsec} produce the same angular distribution. For simplicity, we consider here the scalar-scalar Lorentz structure. We show in fig.~\ref{fig:SS} the resulting cross section multiplied by the scattering angle (a proxy for the differential cross section $d\sigma$ in the small angle approximation) for a number of different choices of dark matter mass and incident proton energy.
The plots show that there exists a peak angle at which scattering is most likely. We then computed the most-likely scattering angle as a function of the incident proton energy by maximizing $\frac{d\sigma}{d\Omega^*}d\theta^*$ as a function of $\theta^*$, and found that, in the limit of small angles, or high cosmic-ray energy, the data are fit by the simple expression (we show one example of the fit in the right panel of fig.~\ref{fig:SS})
\begin{equation}\label{eq:thetaSS}
    \theta_{S}(m_\chi,E)\simeq 0.8^\circ\ \sqrt{\left(\frac{m_\chi}{\rm GeV}\right)\left(\frac{10^{14}\ {\rm eV}}{E}\right)}.
\end{equation}

Note that the structure of this expression (albeit not its precise numerical value) is in fact not surprising, since it reflects the ``relativistic beaming'' of angles from the center of mass to the laboratory frame: the corresponding Lorentz factor $\gamma=1/\sqrt{1-\beta^2}\sim \sqrt{m_\chi/E_2}$, since $\beta$, the laboratory speed, is $\beta=|\vec p_2|/(E_2+m_\chi)$ and thus $1-\beta^2\simeq m_\chi/E_2$

Using the expression in Eq.~(\ref{eq:thetaSS}), we can calculate the typical angular deflection $
\Delta\theta$ experienced by a CR for a given cross section $\sigma_{\chi p}$, taking into account that the angular direction of each scattering is stochastic, 
\begin{equation}\label{eq:deflectS}
    \Delta\theta\simeq \theta_{S}(m_\chi,E)\sqrt{\frac{\Sigma}{m_\chi}\  \sigma_{\chi p}},
\end{equation}
whereas before $\Sigma$ is the DM column density as above. Note that quite remarkably we find that {\em the deflection angle is independent of the dark matter mass}, since the factor $m_\chi^{-0.5}$ from the $\sqrt{n}$ of the random walk in angles cancels exactly with the $m_\chi^{+0.5}$ in Eq.~(\ref{eq:thetaSS}). Of course, this only holds as long as $n\gg 1$.


\subsection{Vector Lorentz structures}
\begin{figure}
    \centering
    \includegraphics[scale=0.3]{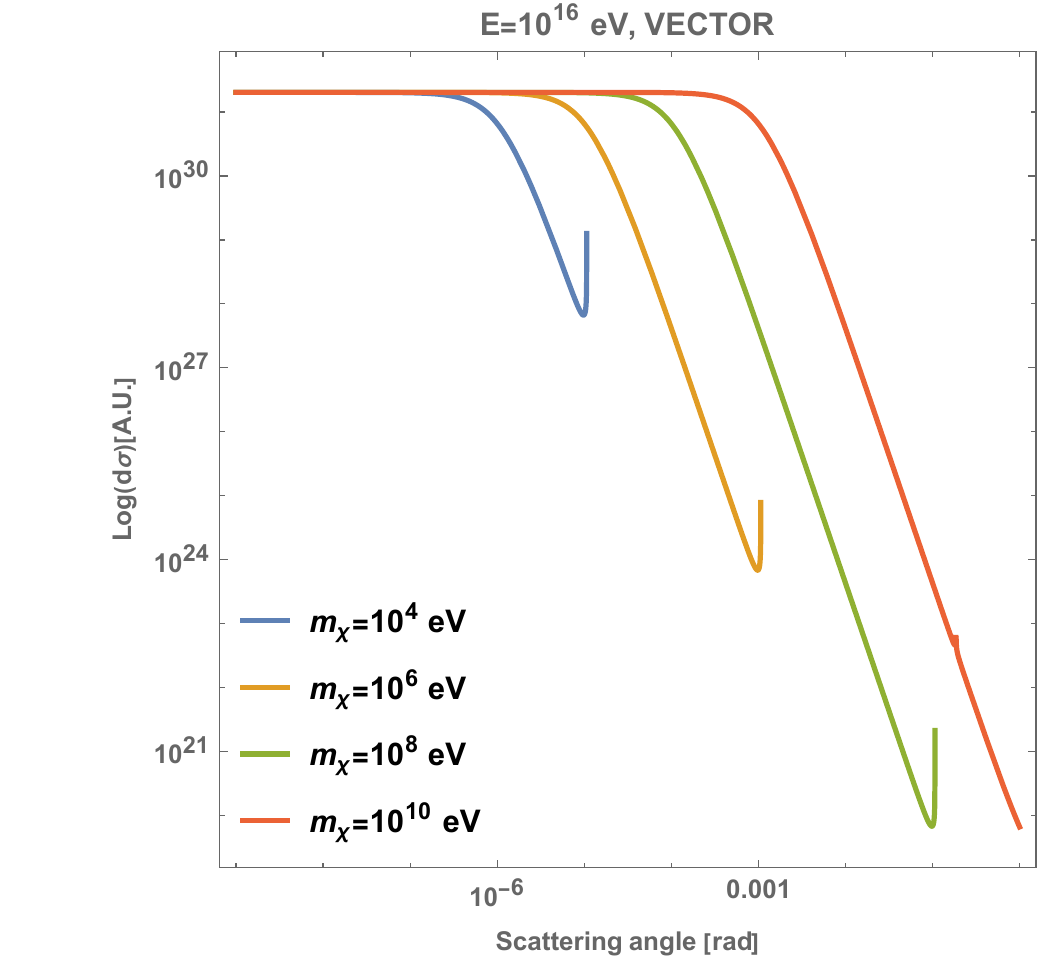}\ \ \includegraphics[scale=0.3]{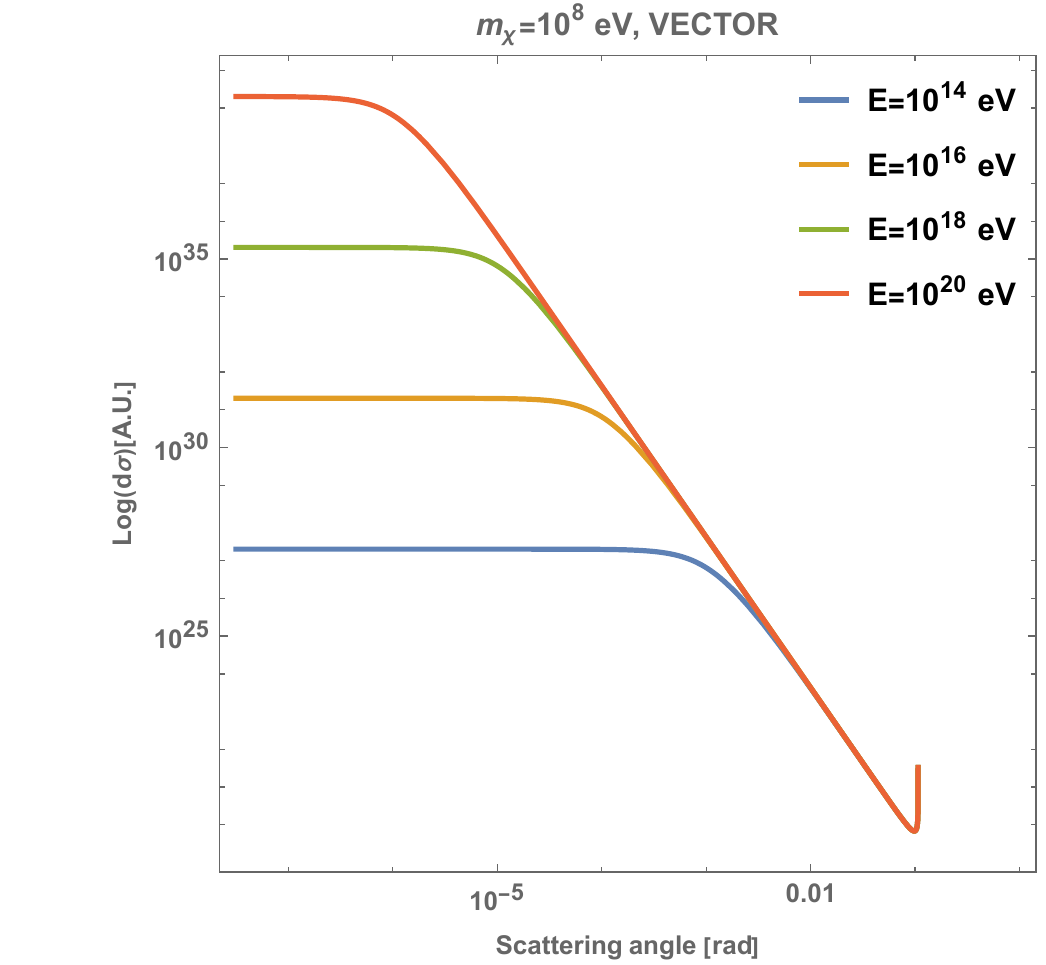}\ \ \includegraphics[scale=0.3]{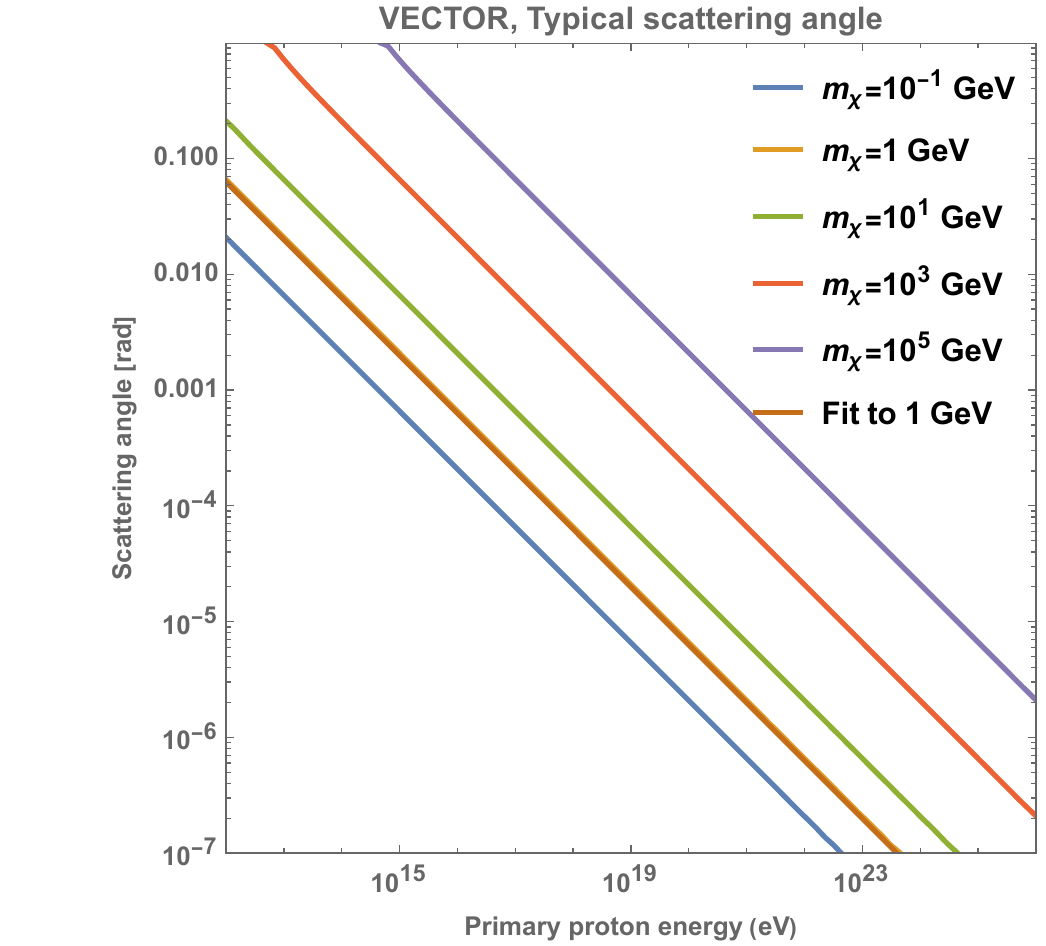}
    \caption{The product of the laboratory cross section and the laboratory scattering angle $\frac{d\sigma}{d\Omega^*}d\theta^*$, as a function of the scattering angle $\theta^*$, for the scalar-scalar cross section, at fixed energy and for different dark matter masses (left), and at fixed dark matter mass, and for different energies (right)}
    \label{fig:AVAV}
\end{figure}

We find that for the spin-dependent cross section for vector structures (for definiteness we utilize the axial-vector-axial-vector structure here), the cross section plateaus at small angles, suddenly dropping at some particular angular value, dependent on the dark matter and on the incident cosmic-ray energy, according to the expression
\begin{equation}\label{eq:thetaAVAV}
    \theta_{V}(m_\chi,E)\simeq 0.36^\circ\ \sqrt{\left(\frac{m_\chi}{\rm GeV}\right)\left(\frac{10^{14}\ {\rm eV}}{E}\right)}.
\end{equation}
The expression, again, albeit with different normalization factors, has the same origin -- relativistic beaming -- as the one relevant for the scalar Lorentz structures. Here, the probability of scattering at small angles is constant, but the average angular dispersion again depends only on the scale set by $\theta_{V}(m_\chi,E)$. Carrying out an identical analysis as before, we find that here again the average angular dispersion is independent of mass, with the angular deflection given as above by
\begin{equation}\label{eq:AVAVangle}
    \Delta\theta\simeq \theta_{V}(m_\chi,E)\sqrt{\frac{\Sigma}{m_\chi}\  \sigma_{\chi p}},
\end{equation}

To within a factor 2, therefore, the results for the vector Lorentz structure cross sections are qualitatively identical to the scalar case shown in fig.~\ref{fig:AVAV}, right. 

\subsection{Predictions for the deflection angle}

\begin{figure}
    \centering
    \includegraphics[scale=0.45]{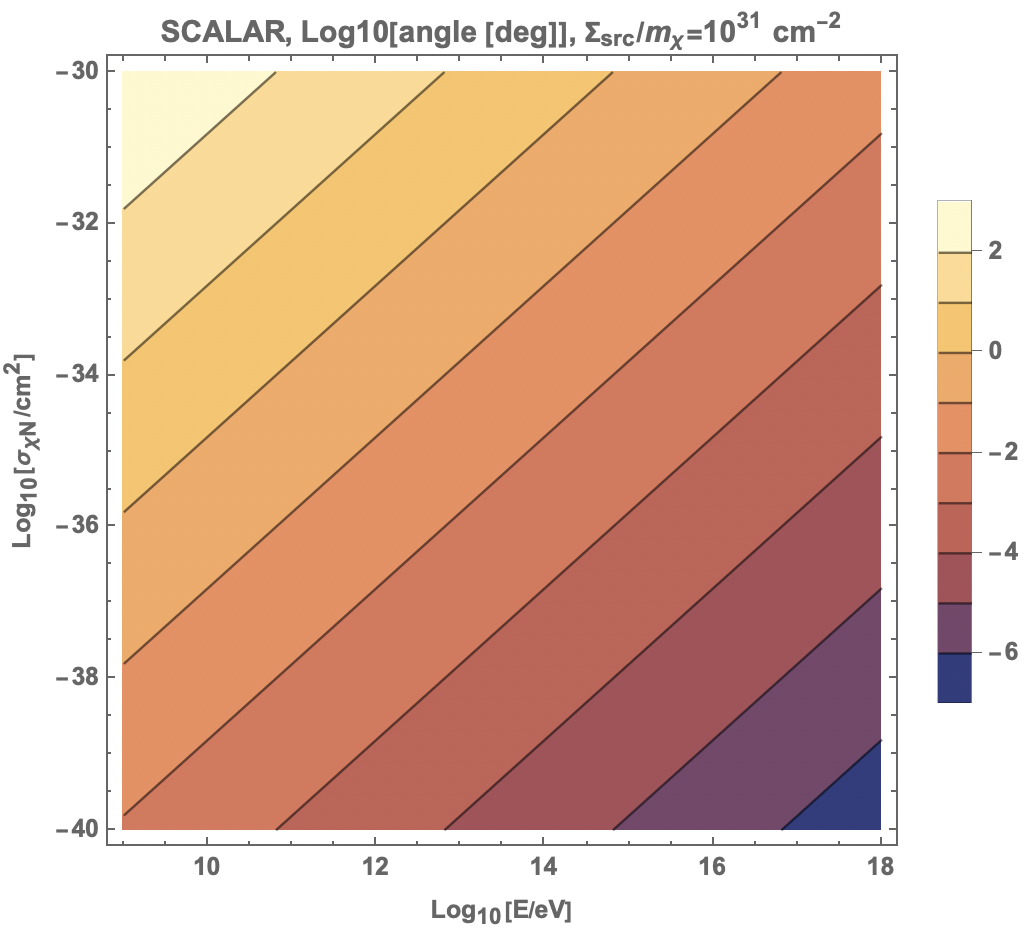}\ \ \includegraphics[scale=0.45]{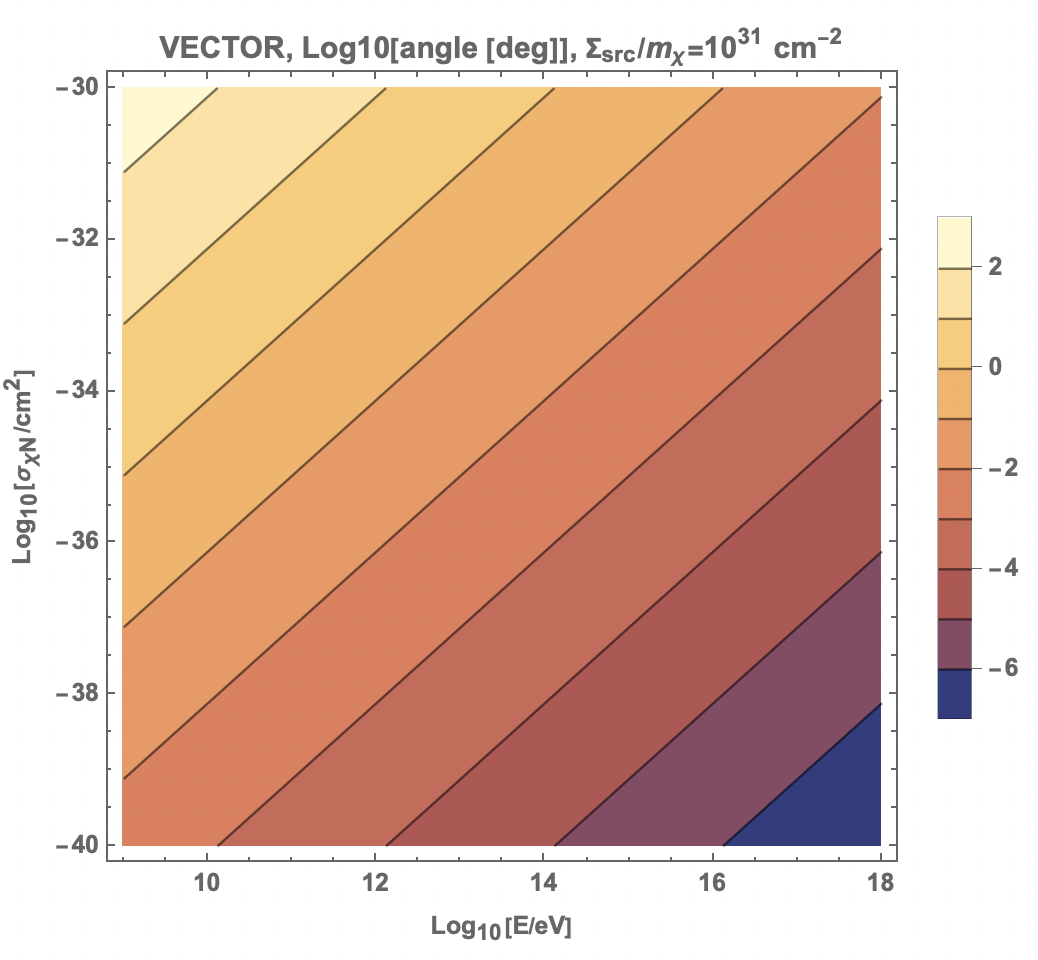}
    \caption{The scattering angle produced by CR-DM scattering, in Log$_{10}$, for an assumed DM column density per unit mass $\Sigma/m_{\rm DM}=10^{31}\ {\rm cm}^{-2}$, for scalar (left) and vector (right) Lorentz structures.}
    \label{fig:angles}
\end{figure}

We gather here the results for the predicted typical deflection angle. We show in fig.~\ref{fig:angles} contours, in degrees and on a logarithmic scale, of the deflection angle for an assumed DM column density per unit mass $\Sigma/m_{\rm DM}=10^{31}\ {\rm cm}^{-2}$, for scalar (left) and vector (right) Lorentz structures as a function of the energy of the impinging CR.

Our results indicate that the deflection of CR trajectories off of DM is most relevant at lower energies and large cross sections, as expected. For the largest-possible cross sections, we find that a deflection comparable to the Pierre Auger observatory's angular resolution, around 2$^\circ$ \cite{PierreAuger:2015eyc}, is possible only for energies below 10$^{15}$ eV in the case of scalar Lorentz structures and 10$^{14}$ eV in the case of vector structures.

\subsection{Finite mass mediator}
\begin{figure}
    \centering
    \includegraphics[scale=0.6]{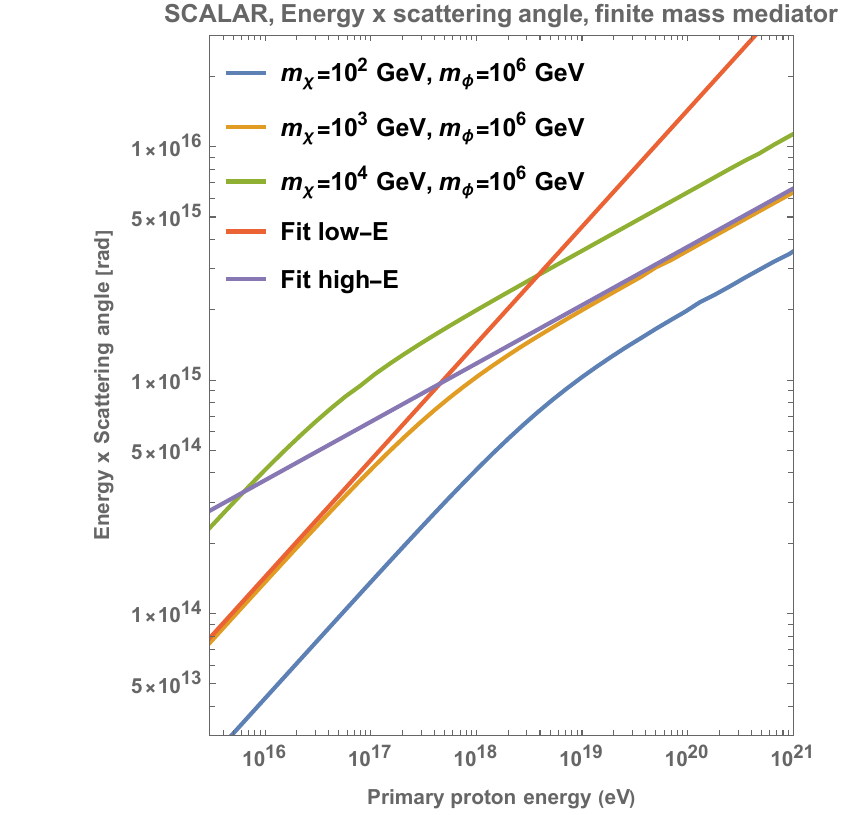}\ \ \includegraphics[scale=0.56]{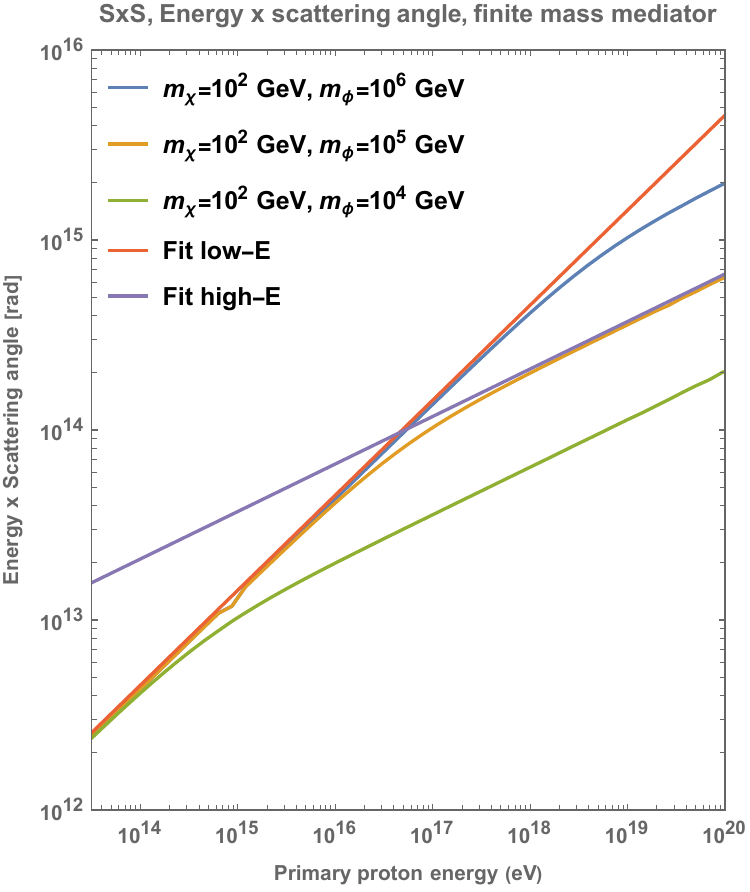}
    \caption{The product of the typical scattering angle times the primary nucleon energy, as a function of the primary nucleon energy, for the scalar-scalar cross section, for $m_\chi=10^2\ 10^3$ and $10^4$ GeV at $m_\phi=10^6$ GeV (left), and for $m_\phi=10^4,\ 10^5$ and $10^6$ for $m_\chi=10^2$. We also show the fits to the low and high-energy behavior of the scattering angle (times energy).}
    \label{fig:angleMS}
\end{figure}

We find that the most-probable angle dependence on cosmic-ray energy and dark matter mass changes drastically with light mediators, i.e. in the regime where $E_2m_\chi \gg m_\phi^2$. Here, we find that, for cross sections with scalar Lorentz structures
\begin{equation}\label{eq:MSH}
    \theta_{\rm MSH}(E_2,m_\chi,m_\phi)\simeq 12^\circ \left(\frac{m_\chi}{\rm GeV}\right)^{1/4}\left(\frac{E_2}{10^{14}\ {\rm eV}}\right)^{-3/4}\left(\frac{m_\phi}{10^6\ \rm GeV}\right)^{1/2}\quad (E_2m_\chi \gg m_\phi^2),
\end{equation}
(MSH indicating finite-mass-Mediator, Scalar, High-energy) with the same functional dependence as before for $E_2m_\chi \ll m_\phi^2$, as expected:
\begin{equation}\label{eq:MSL}
    \theta_{\rm MSL}(E_2,m_\chi,m_\phi)\propto (m_\phi)^{1/2}(E_2)^{-1}.
\end{equation}
We show in fig.~\ref{fig:angleMS} our results on the scattering angle as a function of the CR energy for a few values of the DM mass $m_\chi$, with $m_\phi=10^6$ GeV in the left panel, and of the mediator mass with $m_\chi=10^2$ GeV in the right panel. The figure also shows the low- and high-energy asymptotic behaviors of the scattering angle as a function of the CR energy given in Eq.~(\ref{eq:MSH}) and (\ref{eq:MSL}).

\begin{figure}
    \centering
    \includegraphics[scale=0.41]{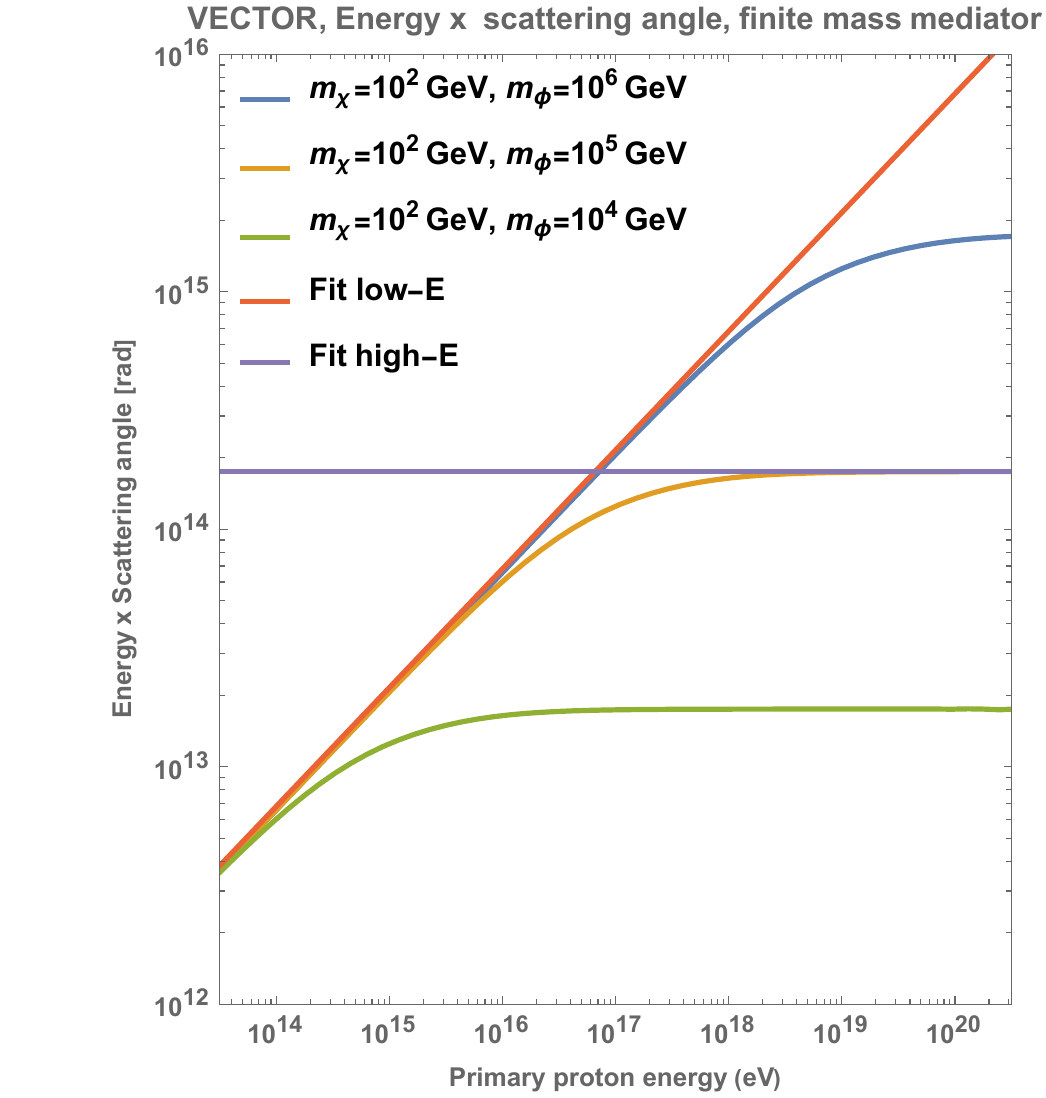}\ \ \includegraphics[scale=0.45]{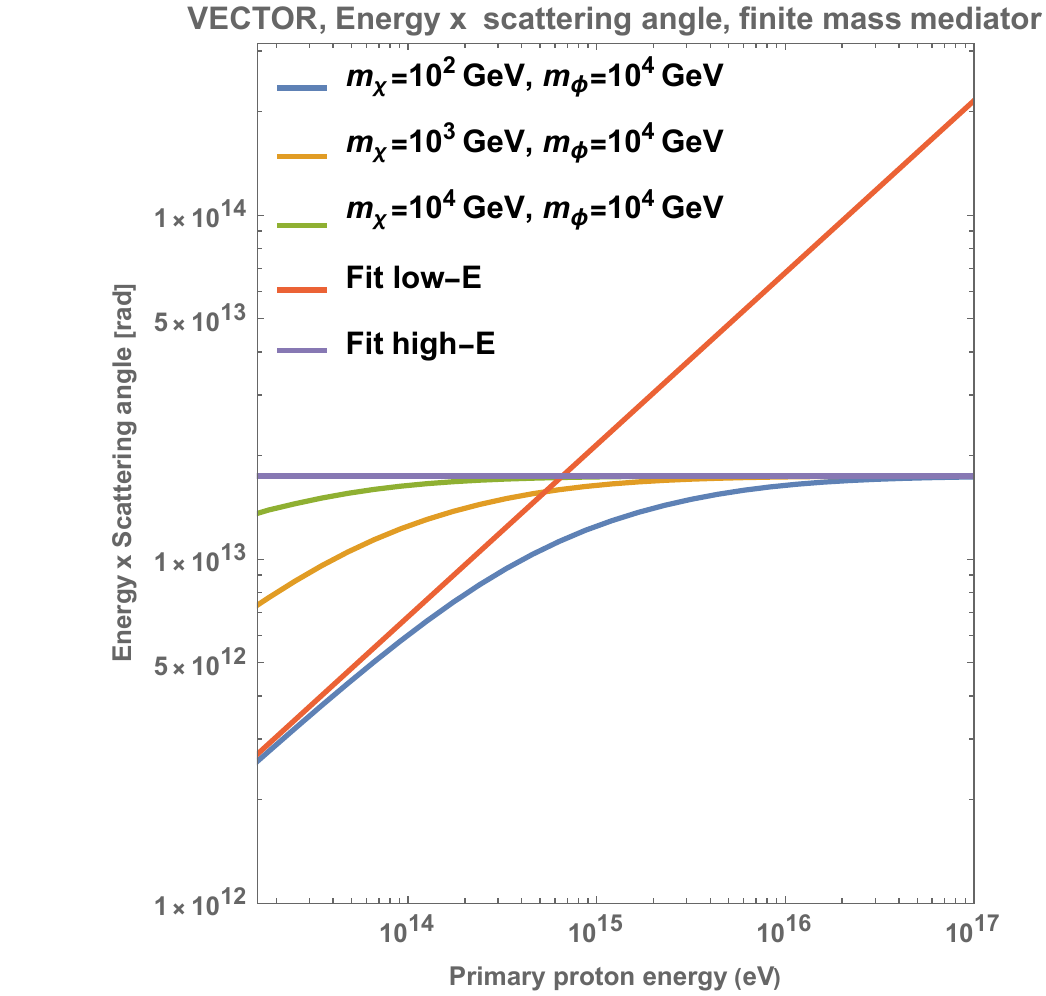}
    \caption{As in fig.~\ref{fig:angleMS} BUT for the vector Lorentz structures.}
    \label{fig:angleMV}
\end{figure}

For cross sections with a vector Lorentz structure, no dependence is found on DM particle mass, in the relevant high-energy range $E_2m_\chi \gg m_\phi^2$, while for $E_2m_\chi \ll m_\phi^2$, where the mediator is heavy and one reduces to the effective theory setup discussed above, we, again, find the same dependence as in Eq.~(\ref{eq:AVAVangle}) above. In the high-energy limit we find instead
\begin{equation}
    \theta_{MV}(E_2,m_\chi,m_\phi)\simeq 10^\circ \left(\frac{m_\phi}{10^4\ {\rm GeV}}\right)^{1/2}\left(\frac{E_2}{10^{14}\ {\rm eV}}\right)^{-1},
\end{equation}
as illustrated in fig.~\ref{fig:angleMV}, again for a variety of mediator and DM masses, and including the asymptotic behavior fits outlined above.

\begin{figure}
    \centering
    \includegraphics[scale=0.45]{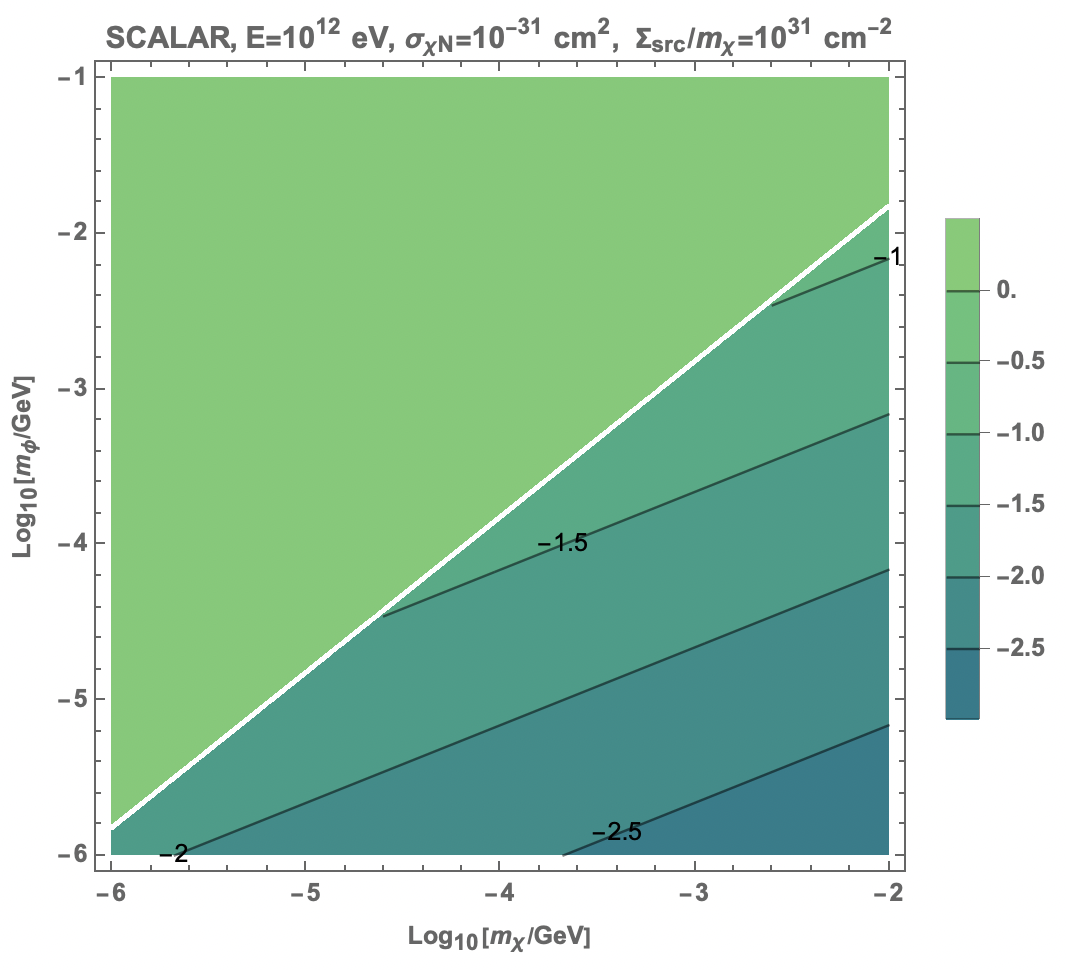}\ \ \includegraphics[scale=0.45]{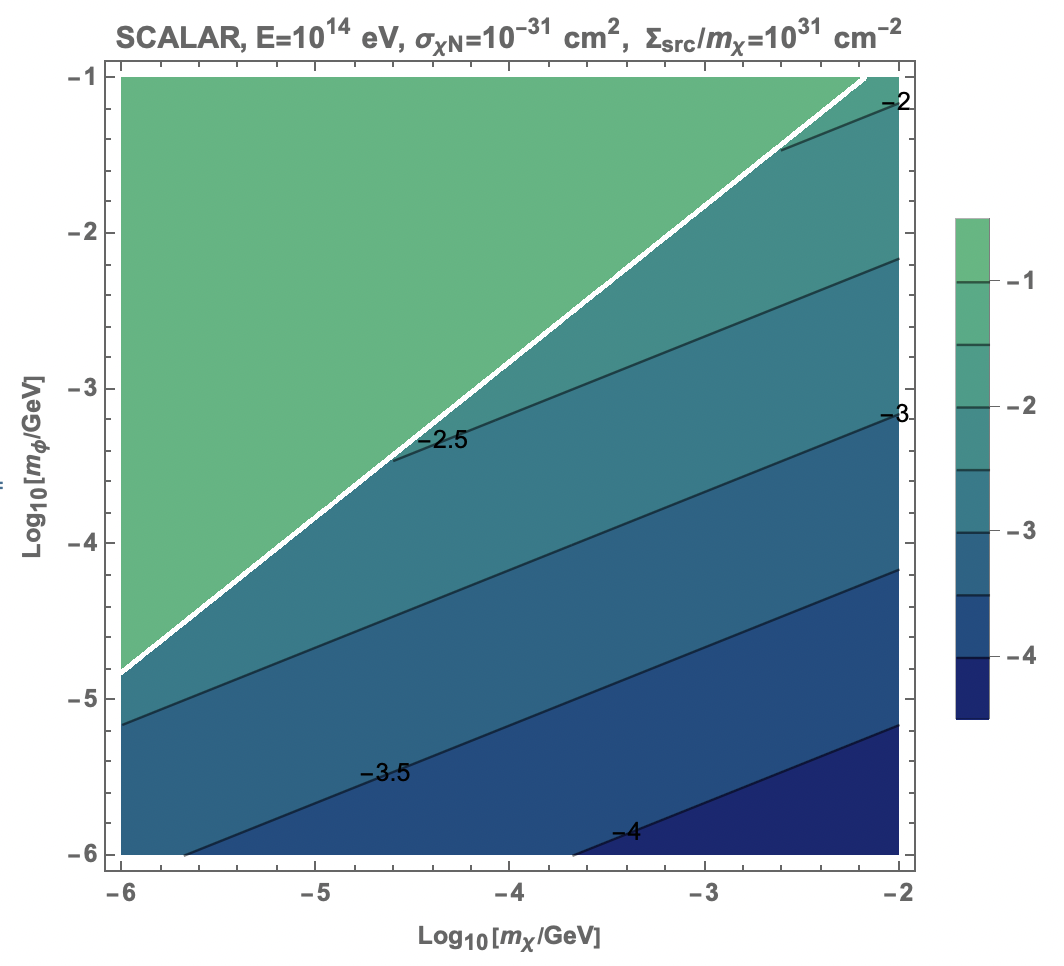}
    \caption{The deflection angle, on a Log$_{10}$ scale, for scalar Lorentz structures, on the plane defined by the DM and mediator masses, at CR energies of$E=10^{12},\ 10^{14}$ eV in the left and right panels, respectively, and for $\sigma_{\chi p}=10^{-31}\ {\rm cm}^2$, $\Sigma/m_{\chi}=10^{31}\ {\rm cm}^{-2}$. }
    \label{fig:SS_ene}
\end{figure}

\begin{figure}
    \centering
    \includegraphics[scale=0.45]{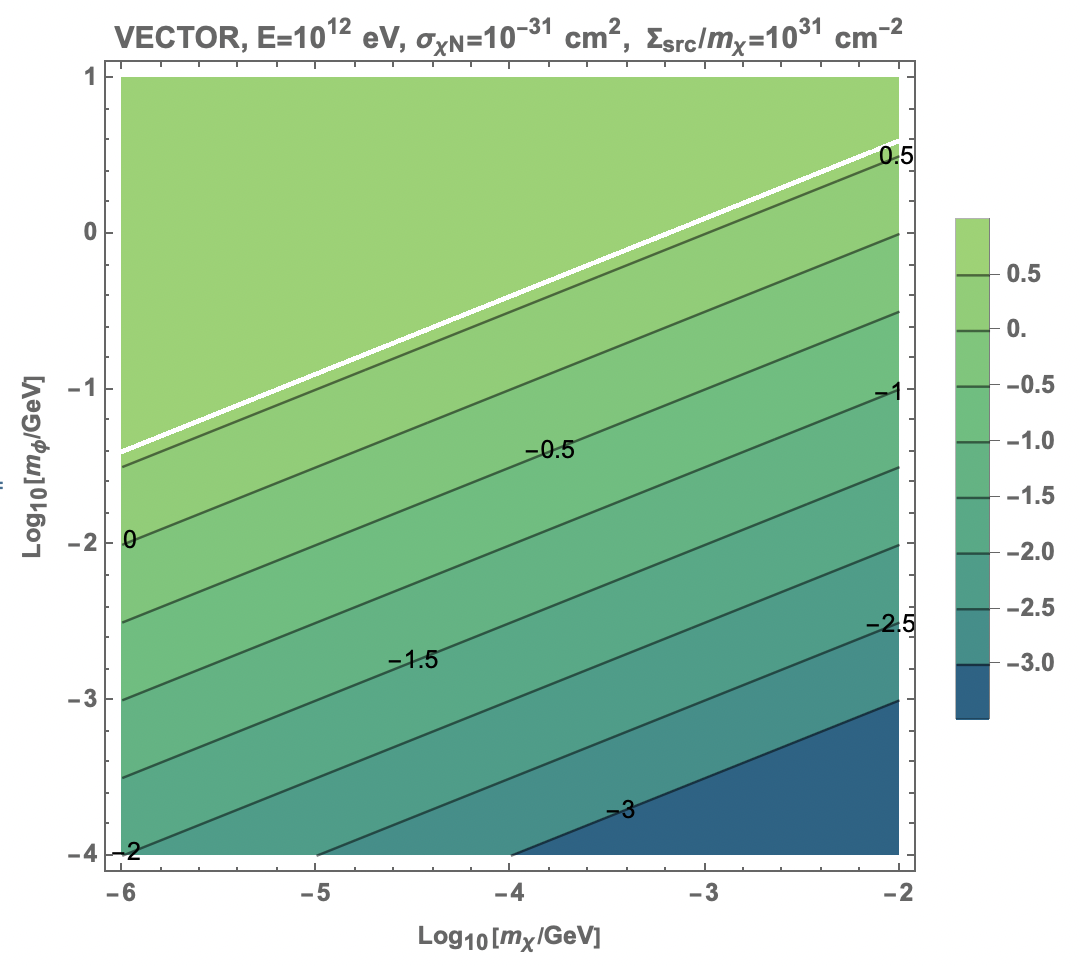}\ \ \includegraphics[scale=0.45]{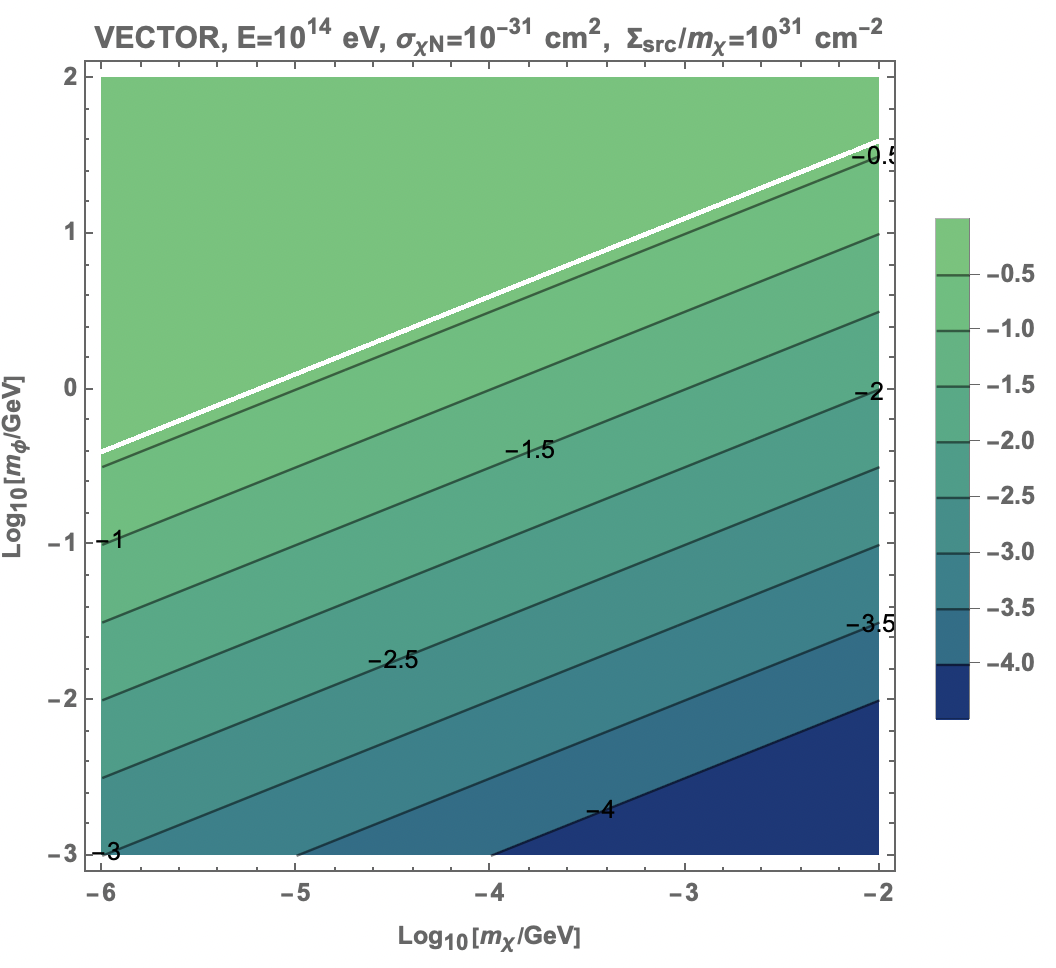}
    \caption{As in fig.~\ref{fig:SS_ene}, but for vector Lorentz structures.}
    \label{fig:VV_ene}
\end{figure}

The addition of a mediator inevitably introduces one more parameter in the theory parameter space. Luckily, the scattering angle scales trivially with the cross section. Thus, at a given energy, we can compute the predicted scattering angle as a function of the DM and mediator mass. This is shown in fig.~\ref{fig:SS_ene} for scalar Lorentz structures and \ref{fig:VV_ene} for vector Lorentz structures; for simplicity, in the figures we set $\sigma_{\chi p}=10^{-31}\ {\rm cm}^2$, $\Sigma/m_{\chi}=10^{31}\ {\rm cm}^{-2}$ and consider two values for the cosmic-ray primary energy, $E=10^{12},\ 10^{14}$ eV in the left and right panels, respectively. 

Generally, and in all cases, we find that the largest scattering angles occur in the regime where $m_\phi$ is large and $m_\chi$ is small, with larger predicted deflection angles at smaller energies.


\section{Discussion and Conclusions} \label{sec:conclusions}
The physical origin of very high-energy cosmic rays remains a mystery and an open question. The ability of ultra-high energy cosmic-ray telescopes to inform this question relies on the control over the deflection of cosmic-ray trajectories as they travel towards the Earth’s atmosphere. While the effect on cosmic-ray trajectories of Galactic and extragalactic magnetic fields is, justifiably, a topic of intense scrutiny, to our knowledge the possible role of very high cosmic-ray scattering off the cosmological dark matter has received, so far, little attention.

The present study intends to clarify the possible role of cosmic-ray-dark matter scattering in deflecting the trajectories of cosmic rays. We considered both an effective field theory and a simplified model description of the relevant proton-dark matter scattering cross section. In both cases, we presented a detailed calculation of both the relevant kinematics, and the cross section, and considered a broad range of Lorentz structures. 

An important element of the computation is to assess the column density of dark matter traversed by the cosmic ray. In doing so, we showed that the column density at the cosmic-ray source is generally dominant over those corresponding to the intergalactic medium and the Galaxy. We argued and demonstrated that, given current constraints on the proton-dark matter cross section at low energies, the most likely regime where dark matter-cosmic ray scattering can be significant is for very light dark matter masses.

We showed that the high-energy cross section, both in the effective field theory description and in the simplified model ``UV-complete’’ description, is dominated by a single term that is identical for each of the scalar-scalar, vector-vector and tensor-tensor Lorentz structures. We also showed that the tensor-tensor case is virtually indistinguishable from the vector-vector case.

The results of the differential angular cross section showed that the scalar case yields a narrow preferred angular scale at which the scattering process occurs, in the ``laboratory’’ frame, as a function of energy, with a simple functional dependence on the dark matter mass and on the cosmic-ray energy. In the case of vector or tensor interactions we similarly found that scattering is equally probable at angles at or below a certain critical angle, which, again, has a simple dependence on the dark matter mass and cosmic-ray energy. We showed that, if not the detailed value of the angle, the functional form of such dependence is to be expected from a well-known relativistic beaming effect.

We then carried out an analysis of the expected deflection on the plane defined by the scattering cross section and the dark matter mass, showing how deflections greater than the Pierre Auger Observatory’s angular resolution are slated to occur at large cross sections and low energies, namely at energies below around 10$^15$ eV or so for the largest-possible cross sections at small dark matter masses.  
Turning to the simplified model description of the cosmic-ray-dark matter interaction, we similarly found that the largest deflections occur at large mediator masses and low dark matter masses.

In the future, it would be interesting to consider specific, well-motivated particle models and compute this effect to ascertain whether in such cases any effect of the type we entertained here is possible.

In conclusion, we showed that indeed cosmic rays at relatively high energies can be deflected by scattering off the cosmological dark matter. We also indirectly shown that given the constraints currently available on the dark matter-proton cross section at low energies, no effect is anticipated, at least in the framework under consideration here, at sufficiently large cosmic-ray energies, specifically in excess of 10$^{15}$ eV or so.

\section*{Acknowledgements} \label{sec:acknowledgements}
   This material is based upon work supported in part by the U.S. Department of Energy grant number de-sc0010107 (SP and MGR).  

\bibliography{bib}


\end{document}